\documentclass[preprint,showkeys,aps,prb]{revtex4}
\usepackage[dvips]{graphicx}
\begin{document}

\title{
From Hard Spheres and Cubes to Nonequilibrium Maps with \\
Thirty-Some Years of Thermostatted Molecular Dynamics.
}

\author{
William Graham Hoover with Carol Griswold Hoover \\
Ruby Valley Research Institute                   \\
Highway Contract 60, Box 601                     \\
Ruby Valley, Nevada 89833                        \\
}

\date{\today}

\keywords
{Statistical Physics, Reversibility, Nonequilibrium, Maps, Information Dimension $D_I$ and $D_{KY}$}

\vspace{0.1cm}

\begin{abstract}
This is our current research perspective on models providing insight into statistical mechanics.
It is necessarily personal, emphasizing our own interest in simulation as it developed from the
National Laboratories' work to the worldwide explosion of computation of today. We contrast the
past and present in atomistic simulations, emphasizing those simple models which best achieve
reproducibility and promote understanding. Few-body models with pair forces have led to today's
``realistic'' simulations with billions of atoms and molecules. Rapid advances in computer
technology have led to change. Theoretical formalisms have largely been replaced by simulations
incorporating ingenious algorithm development. We choose to study particularly simple, yet
relevant, models directed toward understanding general principles.
Simplicity remains a worthy goal, as does relevance. We discuss hard-particle virial series,
melting, thermostatted oscillators with and without heat conduction, chaotic dynamics, fractals,
the connection of Lyapunov spectra to thermodynamics, and finally simple linear maps. Along the way
we mention directions in which additional modelling could provide more clarity and yet more
interesting developments in the future.

\end{abstract}

\maketitle

\section{How This Project Began}

This project began with the loss of a long-time colleague and good friend, Francis Hayin Ree, a
generous and productive gentleman-scholar.  As Bill's nine joint publications with Francis had
appeared in the Journal of Chemical Physics from 1963 to 1968, Bill thought JCP well-suited to a
short publication honoring Francis' memory and work. He  wrote and submitted a memorial piece and
soon received two pieces of news from JCP: The bad news: Memorial Articles were not allowed; The
good news: a ``Perspective'' article touching on Francis' and our own joint work would be welcomed
by the journal. Bill's original Memorial article was promptly published in Computational Methods
in Science and Technology\cite{b39}. We then set to work on this Perspective article just as the
2019 Chinese coronavirus arrived and became established in the United States, providing us with
plenty of time at home in Ruby Valley, for contemplation, research, and writing.

A certain amount of ``Perspective'' is a natural side effect of being part of the research scene for
60 years.  There is plenty of opportunity to distinguish useful from useless work and to reflect upon
the difference.  Two things stand out, with a third lurking in the background.  First of all useful
work is reproducible.  In fact the reproduction of science is in itself extremely satisfying. Being
able to follow a trail gives one confidence in blazing his own, leaving sufficient clues so that his
fellows can follow suit.  A second characteristic of useful work is simplicity as recommended by
Occam and Thoreau.  Simplicity saves time in the assessment and reproduction and generalization and
also indicates a kind regard for one's fellows. When Bill read Zwanzig's claim that cubes could bound
results for spheres\cite{b76} it was relatively easy to see that the ``bound'' was mistaken, as Zwanzig's
ideas were expressed clearly. When Bill read Nos\'e's discussion of canonical dynamics\cite{b50,b51} he
entered into a struggle to understand ideas which were more formal than useful and though stimulating,
were in the end too complex in their presentation.  Simplicity through the exploration of simple models
is the key to understanding and utility.

By studying the simple harmonic oscillator\cite{b24} Bill began to see not only the limited usefulness of
Nos\'e's original work but also the compelling vision which had motivated him.  Simplicity can steer us
toward a third research benefit, modular thinking and working, making the repeated use of integrators,
graphics, text software, and the underlying ideas that motivated their development.  It is a clich\'e
that simplicity can be carried too far.

Keeping up with the literature is certainly useful, and nearly essential. At the frontier this involves
realtime communication with those who are generating the work.  It would be very useful if those
journals which are not free to read would sunset their prohibitions after a reasonable time (a year or
two ?) so that those without access to institutional accounts or libraries could more easily contribute
to the research effort. 

\section{Hard-Particle Legacies of Gibbs, Tonks, and the Mayers}

\subsection{Becoming Familiar with Willard Gibbs' Statistical Mechanics}

Bill's scientific journey began with Chemistry from Louise ``Quiz'' Stull at Washington's Woodrow Wilson
High, and continued with Physical Chemistry from Luke Steiner at Oberlin College, with a mix of Gibbs'
Statistical Mechanics and kinetic theory from Stuart Rice at MIT and from Bill's thesis advisor, Andy De
Rocco at the University of Michigan. This necessary, welcome, and exciting scientific groundwork was topped
off by George Uhlenbeck's Boltzmann Equation course at Ann Arbor. Finally Bill was up-to-speed as a Chemical
Physics graduate student, vintage 1958-1961. The John Kirkwood Memorial issue of the Journal of Chemical
Physics (November 1960) and the Mayers' 1940 {\it Statistical Mechanics} text were required reading at
that time. The Mayers' work was concentrated on the virial expansion of the compressibility factor in powers
of the number density $\rho \equiv (N/V)$,
$$                                                                                                                                                                
\frac{PV}{NkT}=1+B_2\rho^1+B_3\rho^2+B_4\rho^3+B_5\rho^4+B_6\rho^5+B_7\rho^6+\dots \ .                                                                            
$$
Exactly these same coefficients represent the focus of Gibbs' work, his canonical partition function
$Z(N,V,T)= e^{-A/kT}$, where $A(N,V,T)$ is Helmholtz' free energy :
$$
\textstyle{
\ln([\frac{Z{(N,V,T)} }{Z_{\rm ideal}(N,V,T)}]^{1/N})
= - B_2\rho^1-\frac{1}{2}B_3\rho^2-\frac{1}{3}B_4\rho^3-\frac{1}{4}B_5\rho^4
             -\frac{1}{5}B_6\rho^5-\frac{1}{6}B_7\rho^6-\dots \ .
}
$$
Andy De Rocco assured his class ``If you know $Z$ you know everything!'' Here ``everything'' includes
the two derivatives of the free energy, pressure and entropy: $dA = -PdV -SdT$. Evidently on the way
to knowing everything, the evaluation of virial coefficients $\{ \ B_n \ \}$, was worth pursuing. Bill
found that Lewi Tonks' 1936 work served as a good introduction.

\subsection{Lewi Tonks' Hard Rod Model for Fluids and Solids} 
In 1936, prior to the Mayers' work, Lewi Tonks had worked out the pressure-volume-temperature equation of
state for hard rods of unit length\cite{b66}, the one-dimensional analog of hard spheres:
$$                                                                                                                                                                
\textstyle{                                                                                                                                                       
\frac{PV}{NkT} = 1/(1-\rho) = 1 + \rho + \rho^2 + \rho^3 + \rho^4 + \rho^5 + \rho^6 + \dots \ .                                                                   
}                                                                                                                                                                 
$$
This benchmark equation continues to provide useful checks on computational virial-series work for
fluids. It also serves as an introduction to the free-volume model of the solid phase\cite{b21}.  If we imagine an
equally-spaced system of hard rods of unit length then each of them can move a distance $(V/N) - 1$ to the
left or to the right, giving a ``free volume'' estimate of the space accessible to a typical particle,
$v_f = 2(V/N)- 2$. For hard parallel squares and cubes an analogous approximation, using the square and
simple cubic lattice structures, gives:
$$                                                                                                                                                                
\sqrt[2]{v_f} = 2(\sqrt[2]{V/N} - 1) \ [ \ {\rm squares} \ ] \ ; \
\sqrt[3]{v_f} = 2(\sqrt[3]{V/N} - 1)\ [ \ {\rm cubes} \ ] \ .                                                      
$$
The corresponding values of the pressure constitute the ``free-volume'' equation of state for the
``compressibility factor'' $PV/NkT \equiv Z$ for hard rods, squares, and cubes:
$$
Z = (1-\sqrt[1]{\rho})^{-1} \ [ \ {\rm Rods} \ ] \ ; \
Z = (1-\sqrt[2]{\rho})^{-1} \ [ \ {\rm Squares} \ ] \ ; \
Z = (1-\sqrt[3]{\rho})^{-1} \ [ \ {\rm Cubes} \ ] \ .
$$
Finite systems of squares\cite{b60} and cubes have been shown to follow this equation of state precisely at
high density\cite{b19}.

The analogous approximations for solid-phase disks and spheres agree with computer simulations with
deviations of the order of unity. Tonks' hard rod model and his paper with its exaggerated title,
``The Complete Equation of State of One, Two and Three-Dimensional Gases of Hard Elastic Spheres'',
served as a prototype for both kinds of hard-particle phases, the fluid and the solid.  In order to
go beyond Tonks' approximate ideas it is necessary to adopt the Mayers' contribution to statistical
mechanics so as to understand the ``star graphs'' necessary to the computation of the virial
coefficients beyond the four known in the precomputer days.

\subsection{The Mayers' Star Graphs}

The Mayers showed that the $n$th virial coefficient can be written as a sum of integrals represented
by $n$-particle ``star graphs''. The ten four-point star graphs contributing to $B_4$ are shown in
{\bf figure 1}.  At Bill's {\it Alma Mater} of 1958-1961, the University of Michigan, George Ford and
George Uhlenbeck had recently generated a list of all the topologically distinct types of star graphs
up to, and including, the 468 7-point star graphs contributing to $B_7$, the seventh term in the virial
series\cite{b18}. For $n>2$ any $n$-point star graph has at least two independent paths linking any pair of points
in the graph.  The simplest star, the ``ring'' star graph has $n$ links while the ``complete star''
graph has the maximum number of links, $n(n-1)/2$. Each link, or bond, or edge represents a corresponding
Mayer ``f-bond''in an $n$-body integral. The Mayer $f$ represents $e^{-\phi/kT}-1$, where $\phi$ is the
pair potential, and so is equal to either zero or minus one for ``hard particles''. Finally each virial
coefficient is proportional to the summed-up configurational integrals of all these star graphs.

\begin{figure}
\includegraphics[width=1.5in,angle=-90.]{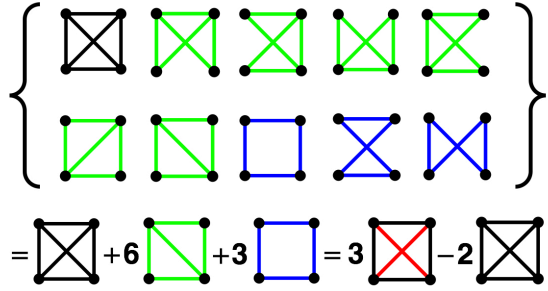}
\caption{
The three types of four-point ``star graphs'' are shown above between braces, $\{ \dots \}$. The
bottom row indicates that the ten summed-up integrals can be reexpressed in terms of the ``complete star''
integral, with six $f$-function links together with a ``Ree-Hoover'' integral with four $f$ functions and
two Boltzmann-factor links (shown in red). For drawings of the star graphs with up to seven points
see reference 18. 
}
\end{figure}

\subsection{Calculating Virial Coefficients for Squares and Cubes at Michigan}

At Michigan, Bill's 1960-1961 thesis work on the virial series for squares and cubes was motivated
by an incautious remark of Bob Zwanzig in the 1956 Journal of Chemical Physics\cite{b76}:
\begin{quote}
``Upper and lower bounds on the virial coefficients of gases composed of rigid                                                                                    
circles or rigid spheres are obtained in  terms of the virial coefficients for                                                                                    
the parallel squares and parallel cubes.''
\end{quote}
To investigate this claim Bill set about generalizing Tonks' hard-rod work to parallel hard
squares and cubes, using the Michigan Algorithmic Decoder ``MAD'' programming language, much like
today's FORTRAN.

The $n$th virial coefficient was expressed by the Mayers as the sum of all the $n$-point star-graph
integrals of the type shown for $n=4$  in {\bf figure 1}. Of the 64 4-point graphs only ten contribute
to the virial coeficient $B_4$. The other 54 four-point graphs do not. The
number of topologically distinct types of star graphs, one three-point graph, three four-point types,
10 five-point, 56 six-point, and 468 seven-point types rapidly approaches $2^{n(n-1)/2}/n!$ for large
$n$. The $n$ or more bonds holding the points of the graph together represent Mayer $f$ functions,
$e^{-\phi/kT} - 1$ where $\phi$ is the pair potential, $k$ is Boltzmann's constant, and $T$ is the
temperature.

\subsection{Virial Coefficients Using Ree-Hoover Graphs\cite{b59}}

{\bf figure 1} showed the Mayers' ten four-point star graphs in the top two rows.  Combining those with
identical integrals, identified by using the same color for the links, gives the first three entries
in the bottom row.  Those three types of four-particle integrals are reduced to just two Ree-Hoover
graphs by introducing the identity $1 \equiv e^{-\phi/kT} - f$. The Mayers' $f$ functions,
$\{ \ e^{-\phi/kT} - 1 \ \}$ correspond to all of the black, blue, and green lines in the figure. The
Boltzmann factors used in the Ree-Hoover graph representation of the integrals (at bottom right) are
the two links shown in red. Of all the Ree-Hoover graphs only the complete-star integral, with its
$n(n-1)/2$ $f$-links, is nonzero when applied to Tonks' hard-rod problem.

The integrals corresponding to the graphs can be evaluated numerically, but there are so many of them
that the enumeration is instead the most time-consuming aspect of the computations. For hard cubes or
spheres or parallel squares and disks the integrands are all $\pm 1$ whenever linked particles overlap,
and zero otherwise. Bill was able to program the evaluation of all the integral types for squares and
cubes, including up to seven points. Summing the star integrals he soon found that both $B_6$ and $B_7$
were negative for hard cubes\cite{b18}. This was quite a surprise, as hard-particle collisions
necessarily make only positive contributions to the pressure. Because both $B_6$ and $B_7$ were thought
to be positive for hard spheres (as Francis Ree and Bill confirmed precisely in 1966), these
discrepancies implied that Zwanzig was wrong. A quick phonecall from Andy De Rocco in Ann Arbor to Bob
Zwanzig in College Park confirmed this conclusion, certainly a ``rare event'', exciting for a youngish
graduate student to discover. It was time for Bill to write his thesis, completed in August 1961.

\subsection{Calculating Virial Coefficients for Disks and Spheres at Livermore}

After a post-doctoral year working on integral equations and their approximate virial series at Duke,
Bill was hired at the Lawrence Radiation Laboratory by Berni Alder and soon began working closely with
a contemporary researcher from Korea, Francis Hayin Ree. Both Berni and Edward Teller provided support
for this work. Francis and Bill introduced ``Ree-Hoover'' graphs in a reformulation of the Mayers' work\cite{b59}.
By introducing $e^{-\phi/kT}$ links as well as $f = e^{-\phi/kT} - 1$ links into the star graphs, they
reformulated the Mayers' expressions, reducing the 468 seven-point star integrals to 171 simpler ones,
each with 21 links of the two different types, $f$ and $e^{-\phi/kT}$. As shown in detail in {\bf figure 1}
the fourth virial coefficient requires only two integral types rather than three, one with 6 $f$s, the
other with four $f$s and two $e^{-\phi/kT}$ links.  Francis and Bill were able to evaluate the integrals
giving $B_6$ and $B_7$ for disks and spheres in 1968\cite{b20}. They were able to estimate the densities
and pressure of the hard-disk and hard-sphere melting and freezing transitions by Pad\'e-approximant
extrapolation.

\subsection{Modern Calculations of Virial Coefficients}
Some 40 years later Nathan Clisby and Barry McCoy used the Ree-Hoover idea to pursue the eighth, ninth, and tenth
coefficients for multi-dimensional hard spheres\cite{b9}, and found that the 9,743,542 ten-point Mayer $f$ graphs
could be expressed in terms of just 4,980,756 ``Ree-Hoover'' graphs with two kinds of links, $f$ and
$e^{-\phi/kT}$. Clisby and McCoy discovered that all the virial coefficients through the tenth are positive
for one, two, three, and four-dimensional hard spheres but that the fourth coefficient becomes negative
for the first time for eight-dimensional spheres, hard particles whose cartesian description requires eight
orthogonal directions in eight-dimensional space rather than the simpler two dimensions of disks and
three of cubes. The sign change is easily understood, in retrospect.  As dimensionality
increases the $n$-link ring integral comes to dominate each exponent. For hard particles, with all the
ring-integral $f$ links negative, $(-1)$, this causes the virial coefficients after $B_2$ to alternate in
sign, with $B_4$, $B_6$, $B_8$, \dots necessarily negative and $B_3$, $B_5$, \dots positive, as indicated in
Clisby and McCoy's work.

In seven more years Richard Wheatley evaluated the eleventh and twelfth coefficients for three-dimensional
hard spheres\cite{b70}. At the freezing density, about two-thirds of close packing, the hard-sphere compressibility
factor is about 12.5. The 12-term virial series sum at that density is 12.39.  Of all the statistical-mechanical
topics which we have studied over the past sixty years the virial series, for both hard-particle and
``realistic'' models, has progressed the most\cite{b71}. In 2014, Zhang and Pettitt\cite{b75} published the first 64
coefficients in the series for multidimensional hyperspheres of as many as 100 dimensions!

We worked on two-dimensional squares and three-dimensional cubes with Marcus Bannerman in 2009. The
three of us found that parallel squares likely have a second-order (discontinuity in slope) phase
transition at a density near 0.793 and that cubes behave even more smoothly.  So far these results
have no theoretical explanation, though they are well established from the numerical standpoint. Wheatley's
valuable contributions suggest that for hard squares and cubes the truncated virial series is as
reasonable a choice as the approximants.  For both squares and cubes the discrepancies between the
truncated and Pad\'e equations of state increased with the additional knowledge of $B_8$ and $B_9$.

The number and complexity of the star graphs limited the number of computationally feasible terms
to seven in 1961.  By 2013 that number had increased, from seven to twelve, through the efforts of
Nathan Clisby, Barry McCoy, and Richard Wheatley\cite{b9,b70}.

Disks and spheres have already been thoroughly investigated. Now, in 2020, it is high time that the
hard-cubes model be reinvestigated, taking advantage of the tremendous computational progress in both
hardware and algorithms  made since 1961. The mechanism for the hard-disk melting transition was
identified in 1962. That and the melting and freezing transitions' locations are the subjects of the
following subsection.

\subsection{Pad\'e Approximants, A Useful Extrapolation Tool?}

\begin{figure}
\includegraphics[width=2.5 in,angle=-90.]{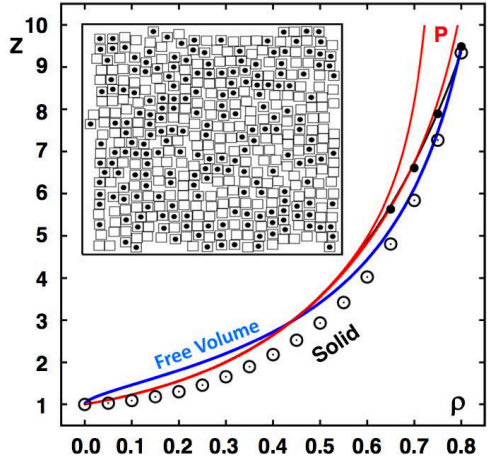}
\caption{
Hard-square equations of state with the compressibility factor $Z=PV/NkT$ shown as a
function of the density relative to close packing $\rho$. ``Solid'' data are from single-occupancy
cell simulations, distinct from the free-volume solid-phase approximation, $(1-\sqrt{\rho})^{-1}$,
shown in blue.
The four filled circles are from 1600-square simulations using 1000 collisions per particle.  The two Pad\'e
approximants (right to left and shown in red) indicate the change from using seven virial coefficients
to using nine.  These two approximants are ratios of cubics and quartics in the density, respectively.
The inset shows a typical configuration of 400 squares at $\rho = 2/3$ where the dotted particles
originally occupied the even-numbered rows.  Marcus Bannerman's minute-long YouTube presentation of
``Parallel Hard Cubes'' is worth well more than a thousand words.
}
\end{figure}

The steady progress in finding virial coefficients suggests a program of extrapolation and checking,
much like the economists' and climate scientists' efforts to understand a country's financial and
environmental futures.  For virial coefficients a systematic nineteenth-century extrapolation
procedure, used by Henri Pad\'e and Georg Frobenius, can be used to generate predicted sums of the
entire virial series. Here is an example Pad\'e approximation, extending the truncated five-term
virial series for hard parallel squares to a closed form, a ratio of quadratics in the density.
$$
\frac{PV}{NkT} \simeq
\frac{1 + N_1\rho + N_2\rho^2}{1 + D_1\rho + D_2\rho^2} \simeq
1 + B_2\rho + B_3\rho^2 + B_4\rho^3 + B_5\rho^4  \ ,
$$
the unknowns in the numerator $\{ \ N_i \ \}$ and denominator $\{ \ D_i \ \}$ can be evaluated by
equating coefficients of like powers of the density. The resulting system of linear equations is then solved
by matrix inversion. The four coefficients $B_2$ through $B_5$ are enough information to identify two
unknowns each in the numerator and denominator:
\[ \left( \begin{array}{cccc}
1 & 0 & -1   & 0    \\
0 & 1 & -B_2 & -1   \\
0 & 0 & -B_3 & -B_2 \\
0 & 0 & -B_4 & -B_3 \\
\end{array} \right)
\left( \begin{array}{c}
N_1\\
N_2\\
D_1\\
D_2\\
\end{array} \right)
=
\left( \begin{array}{c}
B_2\\
B_3\\
B_4\\
B_5\\
\end{array} \right) \ .
\]
The solution of this example problem is :
$$                                                                                                                                                                                  
\frac{1.000000 -0.133335\rho + 0.099999\rho^2}{1.000000 - 2.133335\rho + 1.366669\rho^2} \simeq                                                                                     
1 + B_2\rho + B_3\rho^2 + B_4\rho^3 + B_5\rho^4 \ .                                                                                                                                 
$$

For hard parallel squares Hoover and DeRocco found the following results (rounded to six figures after
the decimal point):
$$                                                                                                                                                                                  
\{ \ B_2 = +2.000000 \ ; \ B_3 = +3.000000 \ ; \ B_4 = +3.666667 \ \} \ ;                                                                                                            
$$
$$                                                                                                                                                                                  
\{ \ B_5 = +3.722222 \ ; \ B_6 = +3.025000 \ ; \ B_7 = +1.650648 \ \} \ .                                                                                                            
$$
For cubes they found:
$$                                                                                                                                                                                  
\{ \ B_2 = +4.000000 \ ; \ B_3 = +9.000000 \ ; \ B_4 = +11.333333 \ \} \ ;                                                                                                           
$$
$$                                                                                                                                                                                  
\{ \ B_5 = +3.159722 \ ; \ B_6 = -18.879630 \ ; \ B_7 = -43.505432 \ \} \ .                                                                                                         
$$
In 2020 Wheatley reconsidered the parallel square and cube systems, finding that both $B_8$ and $B_9$ are
negative for squares, the first negative exponents found for that system.  The exponents for cubes
continue to be much larger, indicating a much less useful series in three dimensions than in two :
$$                                                                                                                                                                     \{ \ B_8 = -0.04094 \ ; \ B_9 = - 1.4125 \ ; \ B_{10} = -1.710 \ \}_{\rm squares} \ ;
$$
$$                                                                                                                   
\{ \ B_8 = -37.3010 \ ; \ B_9 = +40.8927 \ ; \ B_{10} = +171.5 \ \}_{\rm cubes} \ .
$$
Cell models describing the motion of a single particle in the fixed field of its neighbors are not just
crude approximations of reality.  For hard particles one can imagine a very light specimen in a system
sampling its cell while its fellows scarcely move.  It is evident that this picture can be made exact
and it would be an excellent thesis project to make this concept rigorous. For more details see the
discussion of hard-disk free volumes in subsection II.k.

\subsection{Melting/Freezing, {\it via} the Correlated Cell Model}
\begin{figure}
\includegraphics[width=1.5 in,angle=-90.]{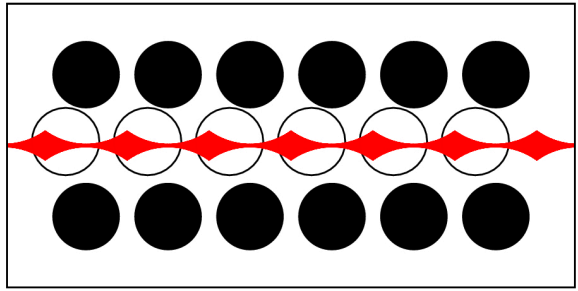}
\caption{
The correlated cell model\cite{b3} describes the (red) free volume available to a row of moving (white)
disks constrained by fixed (black) disks in neighboring rows. At a density three fourths of the
close-packed the free volumes join up, as shown here, and  allow ``hexatic'' sliding, causing a
van der Waals loop in the equation of state.
}
\end{figure}

In the 1950s Berni Alder and Tom Wainwright developed hard-sphere and square-well molecular dynamics
at Livermore\cite{b2} while Bill Wood and Jack Jacobson implemented Metropolis' Monte Carlo algorithm
at Los Alamos\cite{b72}, with the two groups cooperating closely and comparing their results. The
existence of a sharp fluid/solid transition was still controversial in 1957. In January, at a meeting
at the Stevens Institute George Uhlenbeck requested a show-of-hands on the question of belief or
disbelief in a ``transition point'' for hard spheres.  According to the record\cite{b54} he announced
the verdict, ``Even, again!'', recalling a Seattle vote on the same question the previous Fall.  By
mid-August 1957 the existence of a first-order solid-fluid transition was ``strongly suggested'' by
Wood and Jacobson\cite{b72} and ``strongly indicated'' by Alder and Wainwright\cite{b1} in their
near-simultaneous cooperative publications. Memories of this valuable four-man collaboration evidently
faded somewhat over time\cite{b5,b74}.

\subsection{The Hard-Disk Hexatic Phase}

The situation was then a bit less clear for disks than for spheres. Disks are well understood today\cite{b11}.
Disks exhibit a ``hexatic'' phase prior to melting, a modification of the usual solid with cooperative motion
parallel to the three sets of close-packed rows of particles.  Wood remarked on a circular
``ring-around-the-rosey'' cooperative motion of disks in his voluminous 1963 Los Alamos report\cite{b73}.
By then Alder and Wainwright's movies had shown row-wise hexatic motion as a precursor to melting. They and
Bill developed a simple ``correlated cell model'' incorporating the row-wise motion in a two-disk periodic
cell. See {\bf figures 3 and 4}. The periodic boundary condition provides a geometry equivalent to the motion
of two infinitely-long parallel rows of disks. With one of the two particle rows fixed the free area integral,
$a_f = [ \ \int\int e^{-\phi/kT}dxdy \ ]_{\rm cell}$, available to the other one is:
$$
\textstyle{
a_f=\sqrt{3}D^2 - \frac{1}{2}\sqrt{4D^2-D^4} - 2\arcsin(\sqrt{D^2/4}) \ .
}
$$
Here $D>1$ is the center-to-center separation of the unit-diameter disks in the perfect lattice.
The disk number density, relative to close-packing is $D^{-2}$. At densities greater than three-fourths
of the closest-packed the sliding motion is prevented and there are two additional terms:
$$
\textstyle{
\delta a_f = -\frac{1}{2}\sqrt{12D^2-9D^4}+2\arccos(\sqrt{3D^2/4}) \ .
}
$$
The equation of state for this ``correlated cell model'' follows easily from a numerical
differentiation of the free area $a_f$, $\frac{PV}{NkT}= 1 + [ \ d\ln(a_f)/d\ln(D^2) \ ]$ .
See {\bf figure 4}.

\begin{figure}
\includegraphics[width=3in,angle=-90.]{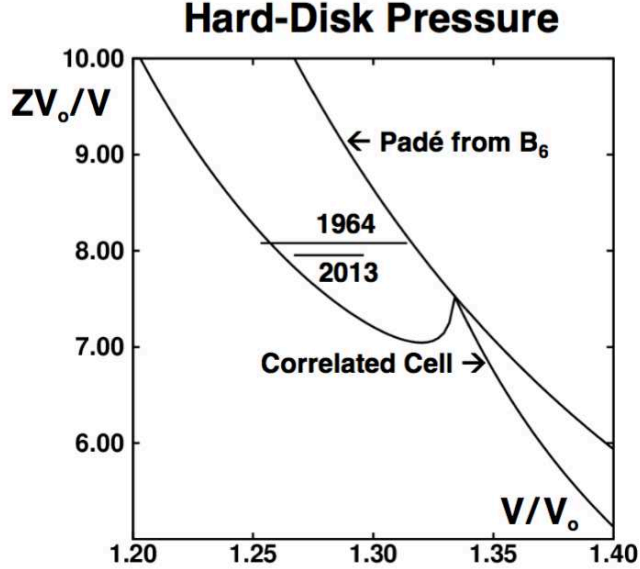}
\caption{
Here we show approximations to $PV_o/NkT$ for hard disks, where $V_o$ is the
close-packed volume. A Pad\'e approximant to the excess entropy, as the ratio of two
quadratics in density, given in reference 19,  was used to estimate the 1964 fluid-solid
tie-line pressure. The accurate 2013 simulations' tie-line pressure comes from reference
11, using $2^{20}$ disks. The single-disk correlated cell model
prediction from reference 3 is surprisingly good. See also reference 4.
}
\end{figure}

  That the sliding mechanism of the correlated-cell model is
reflected in the equation of state can be seen in  the good agreement of the transition
pressure with that measured by a variety of numerical methods. Now, more than half a century
after Alder, Jacobson, Wainwright, and Wood's work, with from 12 to 870 particles, the
hard-disk and hard-sphere transitions have been thoroughly investigated\cite{b11,b40}.  Many
algorithms were implemented and compared for system sizes up to  $2^{20}$ particles,
corresponding to system widths of about 1000 disks and 100 spheres. 

\subsection{Single-Speed Molecular Dynamics for Squares and Cubes}

Although Zwanzig's hard parallel square and cube models were slower to be understood, the
simplicity of their Cartesian collisions suggested useful algorithms for disks and spheres.
In 2009 Marcus Bannerman invented and implemented a special square and cube dynamics in which
every particle's Cartesian velocity components were $\pm 1$. Collisions left the overall
velocity distribution unchanged\cite{b32}.  This same idea, reminiscent of the Ehrenfests'
``Wind-Tree'' model, can be, and has been, applied
to disks and spheres. By 2015 Krauth and his several coworkers had applied a similar
mock dynamics to their Monte Carlo simulations\cite{b11}. Displacement moves of single
disks and spheres were made in the two or three Cartesian directions, with each move
terminating in a mock Cartesian collision.  Just as in the hard parallel simulations these
successive displacements could be continued in Cartesian fashion without altering the
longtime-averaged configurational distribution needed to determine the pressure.  This mock
collision method has the advantage of determining a collision rate directly without the need
for extrapolating binning data to find the probability density of colliding particles.

In addition to the mechanical pressure-volume diagnosis of melting it is feasible, and
certainly equally convincing, to locate transitions by estimating the entropies of
the coexisting phases.  This approach was undertaken by Bill and Francis Ree in 1968\cite{b20}.
The resulting estimates of the transition pressures were $PV_o/NkT$ = 8.08 for disks and
8.27 for spheres, where the close-packed volumes $V_o/N$ are $\sqrt{3/4}$ and $\sqrt{1/2}$
for particles of unit diameter. This approach was also applied to parallel square and cubes
by Bannerman in 2009, but the relative weakness of the transitions in those two cases
precluded definite results.

\subsection{Communal Entropy, Hard-Particle Free Volumes}

Prior to the computer simulations of the middle 1950s theoretical understanding of manybody
systems was restricted to the Mayers' virial series and a variety of approximate models, of
which the ``cell model'' was particularly physical. The configurational states available to
fluid or solid particles were estimated by models of a typical particle's surroundings. See
{\bf figure 5} for a modern computational realization of this idea for 36 hard disks.
Particle motions are restricted both by cell boundaries, enclosing volumes of $(V/N)$
each, as well as by their interactions with neighboring particles. To the left we see free
volumes available to hard disks in a perfect solid structure.  These approximate the $N$th
root of the configurational partition function. Numerical work shows that the $N$th root is
somewhat larger. To the right a Monte Carlo configuration at the same overall density,
0.64 relative to the close-packed density, provides a distribution of free volumes, each of
them corresponding to the instantaneous states accessible to a light particle confined by
its heavier neighbors.

\begin{figure}
\includegraphics[width=2.5 in,angle=-90.]{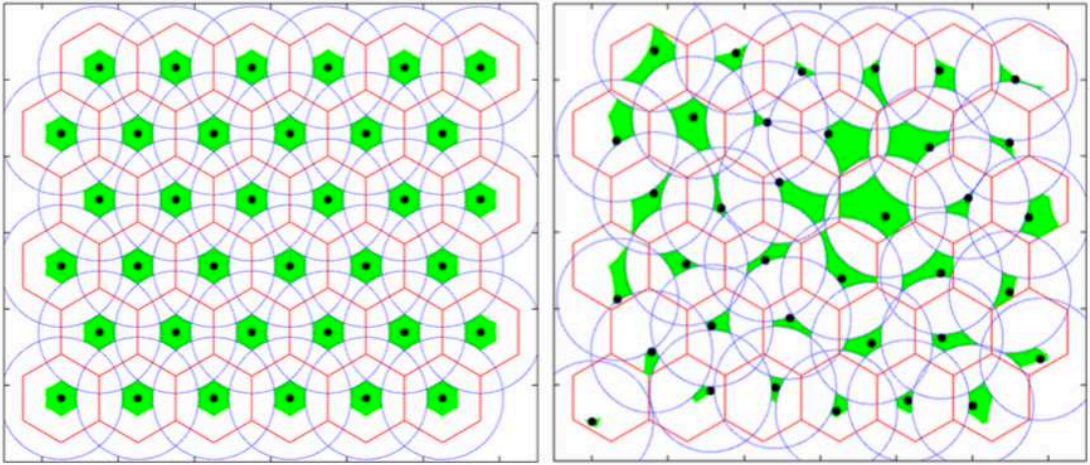}
\caption{
Hard-disk free volumes with periodic boundary conditions from a Monte Carlo
simulation at a density of 0.64 relative to close packing. The volume accessible to each
disk with the others held fixed, is shown in green.  The initial perfect lattice is shown
to the left and a typical fluid configuration appears at the right.  The exclusion disks
drawn around each particle have a radius equal to the disk diameter.
 }
\end{figure}

The cell picture has a defect at low density, predicting a configurational phase volume
smaller than Gibbs' by a factor $e^N$. In 1950, Kirkwood formalized the idea of a ``communal''
entropy as an explanation of the extra factor of e = 2.71828 in the ideal-gas configurational
integral\cite{b43}.
$$
(1/N!)\prod^N_{i=1}\textstyle{[ \ \int_0^vdr_i \ ]} = {\rm limit}_{{N\rightarrow\infty}}
[ \ (V^N)/N! \ ] \simeq  (Ve/N)^N = (ev_f)^N \ .
$$
The division of the $N$-particle integral $V^N$ by $N!$ accounts for the indistinguishability
of the $N$ identical particles. We see in {\bf figure 6} that Kirkwood's communal entropy
dies away slowly, nearly linearly in density, for hard squares, disks, cubes, and spheres.

\subsection{Single-Occupancy Solid Entropy Calculation}

\begin{figure}
\includegraphics[width=2.5 in,angle=-90.]{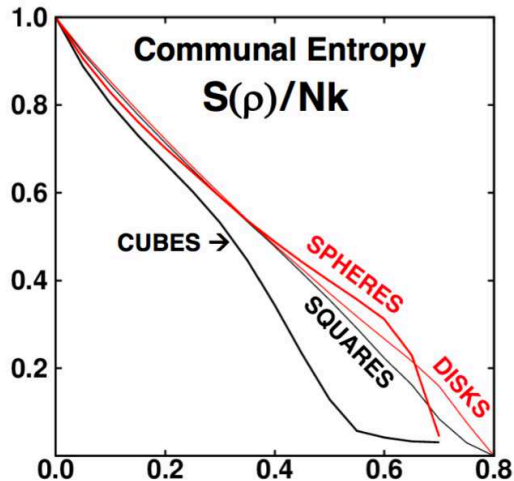}
\caption{
Density dependence of the fluid's entropy relative to the single-occupancy entropy,
Kirkwood's ``communal entropy'' $S_C \equiv S_{\rm Fluid} - S_{\rm SO}$, for four hard-particle
systems\cite{b4,b20,b32}. Presently neither squares nor cubes are thought to exhibit a first-order
freezing transition although Marcus Bannerman is currently reinvestigating this question.
 }
\end{figure}

Francis Ree and Bill adopted Kirkwood's particle-in-cell notion to carry out a precise
numerical evaluation of the solid-phase entropy\cite{b4,b20}. Because the velocity distributions
of the solid and fluid are identical only the configurational entropy determines phase stability.
Kirkwood's idea of confining particles in cells provides a way to extend a ``single-occupancy''
cell model of the solid all the way from close-packing to zero density.  For simplicity we
consider two two-dimensional hard-particle models, hard disks and hard parallel squares.

By confining (the center of) every particle in a manybody system of hard disks or squares to
its own hexagonal or square cell of area $(V/N)$, and integrating the resulting pressure from
low density into the solid phase, the vacancy-free solid's entropy, relative to that of an ideal
gas at the same density and temperature, can be determined with an uncertainty of order $0.01Nk$,
where $k$ is Boltzmann's constant:
$$
\textstyle{
(S_{\rm ideal}-S)/Nk = 1 + \int_o^\rho [ \ \frac{PV}{NkT} - 1 \ ]d\ln(\rho^\prime) \ .
}
$$
We used this approach to determine the coexisting densities for disks (0.761 and 0.798) and
spheres (0.667 and 0.736) relative to close-packing.  For squares the transition density we
determined with Marcus Bannerman in 2009 is 0.793, nearly the same as the melting density for
hard disks. For cubes the number dependence is so large, including many apparent changes in
the sign of the $P(\rho)$ curvature, that no definite  phase-transition density was apparent! Certainly
this is a problem that will eventually be solved through the inexorable exponential advances
in computational capabilities summarized by Moore's Law. With new algorithms additional virial
coefficients and compressibility data for squares and cubes would be a good investment of
computer time. The entropies of simple cubic and brick-wall-structured three-dimensional
crystals are also reasonable and readily accessible research goals.  The entropy associated
with solid-phase vacancies is also readily accessible\cite{b64}.

Bob Zwanzig pointed out that, apart from sign, the Mayers' star integrals in two or three
dimensions are just the squares and cubes of the corresponding one-dimensional hard-rod
integrals\cite{b76}. Additionally, the independence of the spatial distribution to the velocity
distribution means that the molecular dynamics of squares and cubes can be replaced by the
simpler one with the speeds of all particles made identical by choosing each of the Cartesian
velocity components equal to $\pm 1$.

Although treating squares and cubes is relatively simple from the conceptual and computational
standpoints, numerical work reveals considerable complexity.  For example, squares have a smooth
equation of state without any apparent jumps in density as a function of pressure. There does
appear to be  a transition between two distinct phases (fluid and solid), a weak ``second-order''
one, corresponding to a discontinuity in slope rather than magnitude. The situation with respect
to cubes is worse. For ``reasonable'' system sizes such as $10^5$ particles, $P(\rho)$ exhibits
curvature irregularities quite unlike the smooth regular nature of the hard-sphere equation of
state. The large-scale Monte Carlo simulations of 2013 and 2015 show clear van der Waals' loops
for disks and spheres\cite{b11,b40}.  It would be a worthwhile project to follow the number-dependence
of the pressure-volume loops and the entropy, along with additional virial coefficients, as guides to
understanding the square and cube phase diagrams. A rigorous argument for the weighting of free
volumes so as to calculate an average pressure could be a useful part of an interesting thesis.

With a firm grasp on the properties of hard-particle equilibrium systems let us turn next to methods
appropriate to smooth continuous interparticle interactions, the solution of ordinary differential
motion equations, both at and away from equilibrium.

\section{Harmonic-Oscillator Based Models}

\begin{figure}
\includegraphics[width=2.5in,angle=-90.]{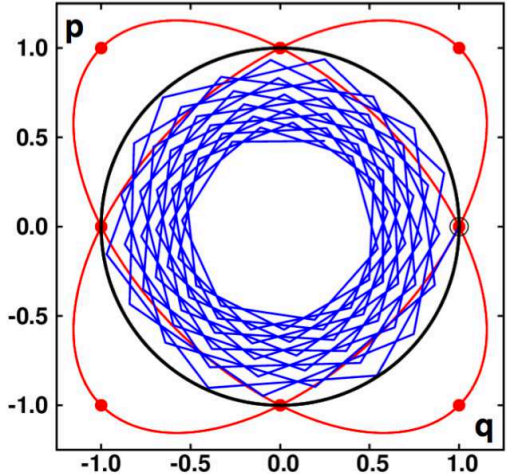}
\caption{
Harmonic oscillator orbits with fourth-order Runge-Kutta $dt = 1$ (blue). Yoshida discovered
a timestep-dependent Hamiltonian, ${\cal H}_{\rm Y} = (q^2 \pm qpdt +p^2)/2$,  which generates
continuuous trajectories that reproduce the St$\o$rmer-Verlet numerical solution. The common
points are shown in red in {\bf figure 7} for the special case $dt=1$. The exact $dt = 0$
solution, $q^2 + p^2 = 1$, is shown in black with the initial point $(q,p)= (0,1)$ emphasized.
}
\end{figure}

Molecular dynamics is the numerical solution of the particle equations of motion, often just
Newton's equations for the motion of point masses, $\{ \ F = ma \ \}$. As acceleration
$a = \dot v = \ddot q$ is the second time derivative of the coordinate, a centered second-difference
approximation, with discrepancy $\frac{dt^2}{12}\stackrel{....}{q}$, can provide the needed algorithm
for a numerical solution. This algorithm's many users have included Loup Verlet, Enrico Fermi, Carl
St$\o$rmer, and Isaac Newton. The one-dimensional harmonic oscillator (with coordinate $q$ and with
both mass and force constant unity) is a useful test case\cite{b67}. For the oscillator there is an
analytic solution:
$$
q(t+dt) = 2q(t) - q(t-dt) - dt^2q(t) \rightarrow q \propto \cos(\omega t) \ .
$$
Assume a periodic solution $q(t) \propto e^{i\omega t}$ and then divide by $2q(t)$.  The result
is :
$$
\textstyle{
\cos(\omega dt) = 1 - \frac{dt^2}{2}\omega^2 + \frac{dt^4}{24}\omega^4 - \dots =
1 - \frac{dt^2}{2} \rightarrow \omega = 1 + \frac{dt^2}{24} + \dots \ .
}
$$
For the extreme example $dt=1$, shown in {\bf figure 7}, the approximate oscillator frequency is
(1/6) revolution per unit of time. This exceeds the exact frequency $(1/2\pi)$ by a factor of
$\pi/3 = 1.0472 \simeq 1 + \frac{1}{24} = 1.041\dot 6$. Typical molecular dynamics timesteps are
ten to one hundred times smaller than atomistic vibrational periods.

{\bf figure 7} shows the periodic six-point oscillator orbit using the finite-difference algorithm
starting at $(q,p) = (1,0)$.  Also shown there is the continuous periodic solution of Yoshida's
Hamiltonian with the finite-difference timestep set equal to unity :
$$
{\cal H}_Y = (q^2 + qpdt + p^2)/2 \longrightarrow
\{ \ \ddot q = -(3/4)q \ ; \ \ddot p = -(3/4)p \ \} \ .
$$
Yoshida showed that the finite-difference solution differs from the exact $dt\rightarrow 0$ limit
by a term of order $dt$.  In the harmonic oscillator example there are no higher-order terms so
that the agreement is exact.

The centered-second-difference St$\o$rmer-Verlet algorithm is perfectly satisfactory for systems
obeying Newtonian or Hamiltonian mechanics, small or large so long as the boundaries and the
energy are fixed.  Simulations at specified values of temperature and pressure required new ideas.
Shuichi Nos\'e furnished them in 1984.

\subsection{Nos\'e's Route to Isothermal Nos\'e-Hoover Dynamics}

In 1984 Shuichi Nos\'e took a giant step toward an ambitious goal, finding isothermal and isobaric
modifications of Hamiltonian mechanics\cite{b50,b51}. We discuss the isothermal case here. Nos\'e wanted to reproduce
Gibbs' canonical distribution with molecular dynamics. He discovered a Hamiltonian, containing what he
called a ``time-scaling'' variable $s$ and its conjugate momentum $p_s$.  For simplicity, which is highly
desirable in view of the complexity of Nos\'e's work, we confine our discussion here to the dynamics of a
one-dimensional harmonic oscillator with mass, force constant, and relaxation time all equal to unity.
In this case Nos\'e's novel Hamiltonian and the motion equations following from it, are
$$
{\cal H}_N = (1/2)[ \ (p/s)^2 + q^2 + p_s^2 + \ln(s^2) \ ] \longrightarrow
$$
$$
\{ \ \dot q = (p/s^2) \ ; \ \dot p = -q \ ; \ \dot s = p_s \ ; \ \dot p_s = (p^2/s^3) - (1/s) \ \} \ .
$$
Nos\'e observed that a logarithmic potential is uniquely capable of providing motion equations
consistent with the canonical distribution. A constant Hamiltonian with the logarithmic potential
implies $s \propto e^{-{\cal H}/kT}$, where Nos\'e termed his $s$ the ``time-scaling variable''.

Unfortunately his time-scaling approach encounters both verbal and computational complexities, with
two sets of variables, ``real'' and ``virtual'', exceptionally ``stiff'' motion equations (as $s$
is often small), and two sets of timesteps. In Nos\'e's words,
\begin{quote}
``The length of the timestep is unequal in the canonical ensemble molecular dynamics method''
[ due to the presence of the time-scaling variable $s$ ].
\end{quote}
Though the idea of adding an additional
thermostatting variable is a good one there are two much simpler routes to Nos\'e's goal. Let us
reverse time in discussing them by first considering Dettmann's approach, from July 1996\cite{b10},
as it is a greatly-simplified version of Nos\'e's. We then take up Hoover's even simpler one, from
1984\cite{b24}, which approaches the canonical-ensemble goal directly, without the unnecessary
distraction of Hamiltonian mechanics.

\subsection{Dettmann's Route to Nos\'e-Hoover Dynamics in 1996}

Bill and Carl Dettmann discussed the need for a Hamiltonian approach to canonical dynamics during a
CECAM meeting at Lyon. Carl discovered that a Hamiltonian similar to Nos\'e's, when set equal to zero
(!), is precisely consistent with the canonical distribution:
$$                                                                                                                                                                              
s{\cal H}_N = {\cal H}_D =  (1/2)[ \ (p^2/s) + sq^2 + sp_s^2 + s\ln(s^2) \ ] \equiv 0 \ .                                                                                       
$$
Choosing ${\cal H}_D = 0$ for the Hamiltonian's value simplifies the time derivative of $p_s$ as follows:
$$
\{ \ \dot q = (p/s) \ ; \ \dot p = -sq \ ; \ \dot s = sp_s \ ; \ \dot p_s =
(1/2)[ \ (p/s)^2 -q^2 - p_s^2 - \ln(s^2) - 2 \ ] = (p/s)^2 - 1 \ \} \ .
$$
Then we can either replace the combination $(p/s)$ by $p$ or equivalently $\dot q$.  With the latter
choice the equation of motion is just the usual Newtonian one with the addition of a time-reversible
friction. Changing the signs of the velocity $\dot q$ and friction coefficient $\zeta$ leaves the
equations of motion unchanged, confirming reversibility.  Any Nos\'e trajectory $q(t)$ can be followed
equally well forward or backward without change despite the ``frictional'' control forces.
$$
\{ \ \ddot q = d(p/s)/dt = (\dot p/s) - (p/s)(\dot s/s) = -q -p_s\dot q \ \} \stackrel{\zeta = p_s}
{\longrightarrow} \{ \ \ddot q = - q - \zeta \dot q \ ; \ \dot \zeta = \dot q^2 - 1 \ \} \ .
$$
Notice that the ``momentum'' $p_s$ becomes the time-reversible friction coefficient $\zeta$ here.

Finally it can be confirmed that the three-dimensional Gaussian distribution function, in either
$(q,p,\zeta)$ or $(q,\dot q,\zeta)$ space, is stationary so that the equations of motion are consistent
with, and preserve, Gibbs' canonical distribution:
$$
f(q,\dot q,\zeta) \propto \exp[ \ -(q^2 + \dot q^2 + \zeta^2)/2 \ ] \ \longrightarrow
 (\partial f/\partial t) = 0 \ .
$$
It is noteworthy that the three-dimensional motion, unlike that in the four-dimensional $(q,p,s,p_s)$
space, is compressible, $(\dot \otimes /\otimes)=-\zeta$, a change that becomes particularly interesting
and important away from equilibrium.  In the end the thermostatted motion equations and the canonical
distribution have both emerged from Hamiltonian mechanics without the need for any time scaling or for
considering the use of unequal timesteps in simulations.  With Dettmann's discovery these complexities
of Nos\'e's ideas are seen to be entirely unnecessary. In fact it is even simpler, and doesn't require
Hamiltonian mechanics at all, to use the phase-space continuity equation to derive a compressible version
of Liouville's Theorem, from which the Nos\'e-Hoover algorithm emerges in a natural way.  This is next. 

\subsection{Hoover's Route to Nos\'e-Hoover Dynamics in 1984}

After lengthy stimulating conversations with Nos\'e in Paris, a few days prior to a meeting in
Orsay, Bill sought to add a friction coefficient $\zeta$ to the Newtonian equations of motion
while constraining the stationary probability density to Gibbs' canonical-ensemble form:
$$
\{ \ f(q,p,\zeta) \propto e^{-(q^2+p^2)/2}e^{-g(\zeta)} \ ; \ \dot p = -q -\zeta p \ \} 
\longleftrightarrow (\partial f/\partial t) \equiv 0 \ .
$$
The continuity equation came in handy here.  For a stationary probability density in the three-dimensional
$(q,p,\zeta)$ space the sum of all six contributions to $(\partial f/\partial t)$ must vanish:
$$
-f[ \ (\partial \dot q/\partial q)+(\partial \dot p/\partial p) 
+(\partial \dot \zeta/\partial \zeta) \ ]
-\dot q(\partial f/\partial q)-\dot p(\partial f/\partial p)
-\dot \zeta(\partial f/\partial \zeta) \equiv 0 \ .
$$
The product form of the distribution function, with $g$ depending only on $\zeta$, turns out to satisfy
the continuity equation in $(q,p,\zeta)$ space, with the control variable $\zeta$ regulating the kinetic
temperature by integral feedback, $\zeta \propto \int^t [ \ p(t^\prime)^2 - 1 \ ]dt^\prime$ ,
$$
0+f\zeta+0+fpq+f(-q-\zeta p)p+f\dot \zeta(dg/d\zeta) \equiv 0 \longrightarrow
$$
$$
g(\zeta) = (\zeta^2/2) \ ; \ \dot \zeta = p^2 - 1 \ .
$$

This Nos\'e-Hoover approach is not at all unique.  Bauer, Bulgac, and Kusnezov discuss the generality
of such thermostatting approaches and note particularly the usefulness of cubic forces in promoting
``ergodicity'', the desirable property of accessing the entire distribution from any initial
condition\cite{b8,b46}. In the case of the oscillator Nos\'e-Hoover dynamics is not at all ergodic.
{\bf figure 8} shows an improved approach toward ergodicity when cubic terms are included.

\begin{figure}
\includegraphics[width=2 in,angle=+90.]{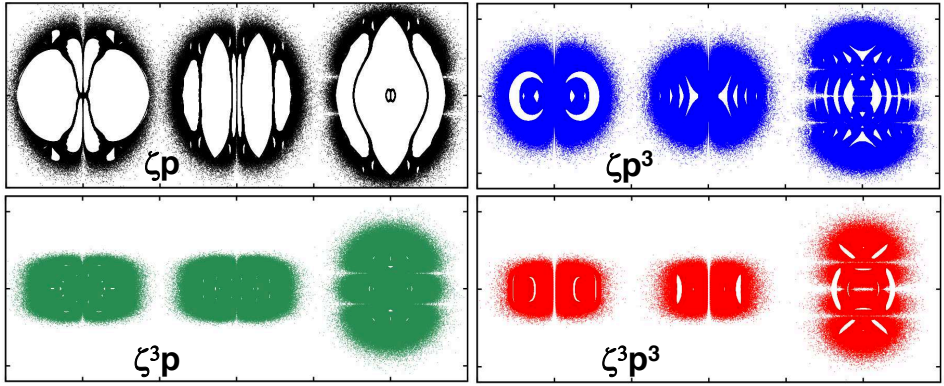}
\caption{
$\{ \ \dot q = p \ ; \ \dot p = -q -\zeta^m p^n \ \} \ {\rm where \ } \dot \zeta = p^2 - 1 {\rm \ for} \                                                                                            n=1 \ {\rm and} \ \dot \zeta = p^4 - 3p^2 \ {\rm for} \ n=3 \ .$
The four simulations each used two billion timesteps with $dt = 0.003$.  From left to right the Poincar\'e
sections plot $(p,\zeta), \ (q,\zeta)$, and $(q,p)$. The tic-mark spacing is five in each of these plots.
}
\end{figure}

This prototypical harmonic oscillator example\cite{b55} only hints at the many applications of Nos\'e-Hoover
dynamics. Though Nos\'e's goal was the dynamical simulation of thermal equilibrium it is evident that
such thermal boundary conditions can also be used to drive and analyze nonequilibrium steady states, the
longstanding and elusive goals of researchers in statistical mechanics\cite{b6,b12,b13,b14,b23}. Nos\'e's
incorporation of
macroscopic temperature into the microscopic motion equations can be generalized to other types of
applications in several ways: [1] to manybody systems at or away from equilibrium; [2] to ergodic
distributions, by using more elaborate thermostats\cite{b8,b46}; and [3] to situations far from
equilibrium\cite{b57}.

In the manybody case it is feasible to apply different thermostat variables $\{ \ \zeta \ \}$ to
different degrees of freedom and to control the local velocities as well as the kinetic temperature.
A straightforward route to ergodicity is to use two control variables rather than one\cite{b28,b48},
constraining both the second and the fourth moments of the velocity distribution:
$$
\{ \ \dot q = p \ ; \ \dot p = -q - \zeta p - \xi p^3 \ ; \ \dot \zeta = (p^2 - 1) \ ; \ \dot \xi =
(p^4 - 3p^2) \ \} \ [ \ {\rm Hoover-Holian \ Oscillator} \ ] \ .  
$$
In subsection III.h we will see detailed numerical evidence that these  motion equations are indeed
ergodic\cite{b35}, filling out a four-dimensional Gaussian probability density,
$$
f(q,p,\zeta,\xi) \propto  e^{-[ \ q^2+p^2+\zeta^2+\xi^2 \ ]/2} \ .
$$

Although these thermostatted algorithms have ``frictional'' control forces in additional to the usual
forces derived from a potential function it is relatively easy to formulate time-reversible
centered-difference algorithms to generate accurate trajectories.  We explored this extension with
Brad Holian and Tony De Groot and describe such an algorithm next for the Nos\'e-Hoover oscillator.

\subsection{Centered Time-Reversible Algorithm for the Nos\'e-Hoover Oscillator}

``Thermostatted'' molecular dynamics such as the Nos\'e-Hoover or Hoover-Holian approach provides
simulations at constant temperature. Let us demonstrate the feasibility of extending an algorithm
of the St$\o$rmer-Verlet type to solve the Nos\'e-Hoover oscillator problem \cite{b17}. The harmonic
oscillator, with the initial conditions $(q,p,\zeta) = (2.21,0,0)$ is an excellent example.  The
underlying differential equations can then be written in either one of two ways :
$$
\{ \ \dot q = p \ ; \ \dot p = -q - \zeta p \ ; \ \dot \zeta = p^2 - 1 \ \}
\ {\rm or, \ equivalently}
$$                                                                                                                                                                              
$$
\{ \ \ddot q = - q - \zeta \dot q \ ; \ \dot \zeta = \dot q^2 - 1 \ \} \
[ \ {\rm Nos\acute{e}-Hoover \ Oscillator} \ ] \ .
$$                                                                                                                                                                              
The second-derivative form suggests a centered-difference analog:                                                                                                               
$$
q_{+dt} = 2q_0 - q_{-dt} + a_0dt^2 - \zeta_0(q_{+dt} - q_{-dt})dt/2 \ ; \
\zeta_{+dt} = \zeta_0 + dt[ \ (q_{+dt}-q_0)^2/dt^2 - 1 \ ] \ .
$$
A convenient initial condition chooses the coordinate $q$ at a turning point and the friction
coefficient zero so that $q_{+dt} = q_{-dt} = q_0 + a_0dt^2/2$. {\bf figure 9} shows an
accurate Runge-Kutta solution of the Nos\'e-Hoover oscillator with initial conditions
$(q,p,\zeta) = (2.21,0,0)$ as a wide line with the centered-difference solution superimposed
as a narrower line.

\begin{figure}
\includegraphics[width=3 in,angle=+90.]{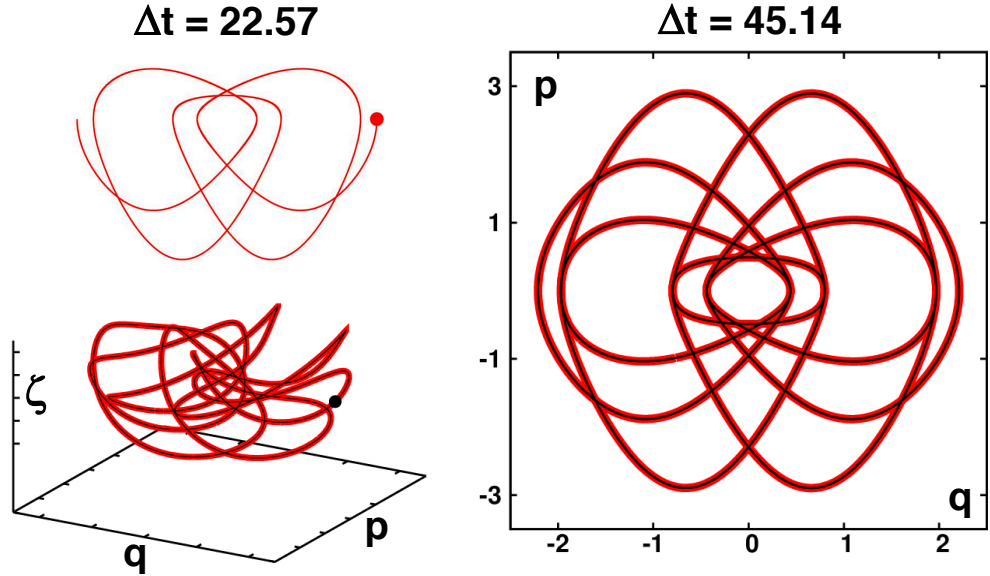}
\caption{
Periodic trajectory of a thermostatted oscillator with
$\{ \ \dot q = p \ ; \ \dot p = -q - \zeta p \ ; \ \dot \zeta = p^2 - 1 \ \}$. The initial
conditions $(q,p,\zeta) = (2.21,0,0)$ are indicated by the red and black dots at the left.
The first half of this periodic Nos\'e-Hoover orbit (upper left) is projected onto the $p(q)$
plane and shows nine trajectory crossings of the three-dimensional orbit (lower left) with
$-2.5 < \zeta < +2.5$. The complete orbit shows 36
crossings. Fourth-order Runge Kutta and centered finite-difference integrator solutions are shown
at the right after one billion timesteps $dt = 0.01$. The centered-difference algorithm remains
close to this orbit for one billion timesteps with $dt = 0.1$ while fourth-order Runge-Kutta loses this
orbit in favor of a knotless loop between $dt = 0.07$ and $dt = 0.08$ in billion-step simulations.
 }
\end{figure}

The centered-difference algorithm has a trio of advantages: stability, simplicity, and reversibility.
The Runge-Kutta algorithm, detailed in the next subsection, is likewise straightforward and simple, though it
lacks longtime stability and time-reversibility.  Beyond these two useful approaches dozens of higher-order
algorithms have been derived and implemented\cite{b34,b37}.  All of them require more storage and more
effort to reproduce. We illustrate the most useful of them here by applying it to a stiff, and therefore
challenging example, the Nos\'e oscillator\cite{b37}, illustrated in {\bf figure 10}.  The corresponding
nonstiff Nos\'e-Hoover oscillator problem is shown there as well. A periodic Nos\'e-Hoover orbit 
appears in {\bf figure 9}.

\subsection{Fourth-Order Runge-Kutta Integrators}

\begin{figure}
\includegraphics[width=3.5 in,angle=+90.]{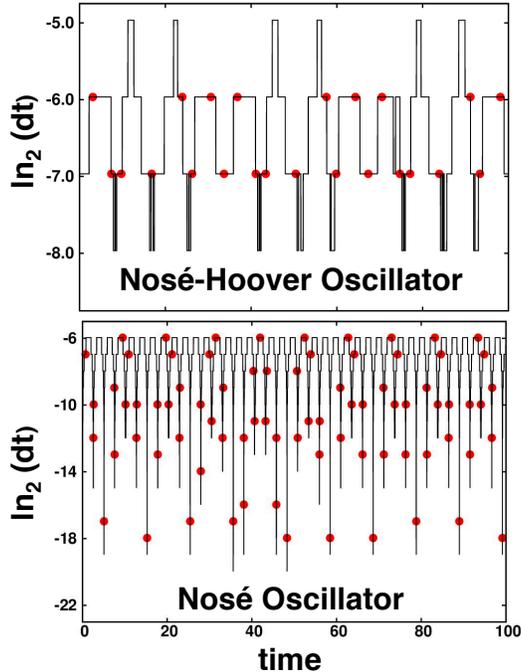}
\caption{
The fourth-order adaptive Runge-Kutta integrator, with an rms error confined
to the interval $10^{-12}$ to $10^{-10}$ shows that the stiffer Nos\'e equations require
timesteps a thousand times smaller than do the Nos\'e-Hoover equations, though both trace
out exactly the same $\{ \ q,p,s,\zeta \ \}$ trajectory (at quite different rates).  The red dots
indicate oscillator turning points where $p = 0$.
 }
\end{figure}

An advantage of Runge-Kutta integrators is that the timestep $dt$ can be changed automatically
in response to stiffness.  By comparing the results of a single timestep to those of two half
timesteps the rms discrepancy can be restricted to a prescribed interval by halving or doubling
$dt$. A comparison of timestep histories for the stiff Nos\'e and nonstiff Nos\'e-Hoover oscillators
is shown in {\bf figure 10} with the details spelled out in reference 37.

Fourth-order Runge-Kutta integrations begin by selecting three
additional points in the vicinity of the initial or current coordinate set of time-dependent
variable, coordinates $q(t)$ and momenta $p(t)$ in the case of Hamiltonian mechanics:
$$
\textstyle{
q_1 = q(t) + \frac{dt}{2}\dot q(t) \ ; \ q_2 = q(t) + \frac{dt}{2}\dot q_1 \ ; \ 
q_3 = q(t) + dt\dot q_2 \ .
}
$$
By expressing the derivatives at each of the four points as Taylor's series it is possible to
choose coefficients giving an averaged derivative such that errors of order $dt$, $dt^2$, $dt^3$,
and $dt^4$ all vanish, with the summed up coefficients equal to unity.  Though this choice is not
unique the symmetric average, ``classic Runge-Kutta'', is nearly always used:
$$
\textstyle{
q(t+dt) = q(t) + \frac{dt}{6}[ \ \dot q(t) + 2\dot q_1 + 2\dot q_2 + \dot q_3 \ ] =
          q(t) + \frac{dt}{6}[ \      p(t) + 2     p_1 + 2     p_2 +      p_3 \ ] \ , 
}
$$
giving an error of order $dt^5/5!$ and completing the algorithm. The algorithm requires a
knowledge of the momentum $p = \dot q$ where $p$ is advanced in just the same fashion as $q$,
using the first-order set of motion equations  $\{ \ \dot q = +p \ ; \ \dot p = -q \ \}$ :
$$
\textstyle{
p(t+dt) = p(t) + \frac{dt}{6}[ \ \dot p(t) + 2\dot p_1 + 2\dot p_2 + \dot p_3 \ ]
        = p(t) - \frac{dt}{6}[ \      q(t) + 2     q_1 + 2     q_2 +      q_3 \ ] \ .
}
$$
Here too the Runge-Kutta algorithm has an analytic solution, but the oscillation is slightly
damped, proportional to $dt^6$.  Although the algorithm is neither reversible nor conservative
it is particularly simple to implement and is a first choice when thermostatted motion equations
are introduced to carry out nonequilibrium simulations. The simplest such system is the Nos\'e-Hoover
Oscillator. In the next subsection we apply the Runge-Kutta technique to a nonequilibrium version
of the harmonic oscillator exposed to a specified temperature gradient $(dT/dq)$, developed by Bill
and Harald Posch in 1997 and revisited in 2014\cite{b57,b62}.

\subsection{Nonequilibrium Nos\'e-Hoover Mechanics with T(q) and Knots}

To begin, we generalize the Nos\'e-Hoover oscillator a bit by introducing the thermostat frequency
parameter $\alpha$ and the temperature $T(q)$, which may be a nonconstant function of location for
which we will choose $T(q) = 1 +\epsilon \tanh(q)$. Throughout we maintain the mass and force
constant of the oscillator equal to unity.
$$
\{ \ \dot q = p \ ; \ \dot p = -q - \zeta p \ ; \dot \zeta = \alpha (p^2 - T) \ \} \
[ \ {\rm Nos\acute{e}-Hoover \ Oscillator} \ ] \ .
$$
When the temperature is constant we have seen that the Nos\'e-Hoover oscillator has a deceptively
simple stationary state.  The stationary probability density $f(q,p,\zeta)$ is a three-dimensional
space-filling Gaussian :
$$
f(q,p,\zeta) \propto e^{-[ \ q^2+p^2+(\zeta^2/\alpha) \ ]/2} \stackrel{{\rm NHO}}{\longleftrightarrow}
$$
$$
(\partial f/\partial t) =
-\partial (f\dot q)/\partial q-\partial (f\dot p)/\partial p-\partial (f\dot \zeta) /\partial \zeta 
\stackrel{{\rm NHO}}{\longleftrightarrow}
$$
$$
0 = -f[ \ -\zeta - qp - p(-q -\zeta p) - (\zeta/\alpha)\alpha(p^2 - 1) \ ] \equiv 0 \ .
$$
In 1985 Runge-Kutta solutions of the oscillator motion provided a stimulating surprise: at most,
the numerical solutions for various thermostat strengths $\alpha$ occupied only small portions of this
Gaussian\cite{b24}. 

For the Nos\'e-Hoover oscillator with $\alpha = 1$ the most common numerical solution type is a torus.
There are infinitely many of these, with each of them centered on a stable one-dimensional periodic orbit
in the three-dimensional $(q,p,\zeta)$ space. In addition to the tori, a definite fraction of the
solutions explores a unique three-dimensional ``chaotic sea''. Motion in that sea is ``Lyapunov unstable'',
meaning that small perturbations grow exponentially with time, an essential characteristic of chaos.

With  $\alpha = 1$ and initial conditions for numerical solutions randomly drawn from the Gaussian stationary
distribution, about six percent of the solutions trace out the three-dimensional chaotic sea\cite{b53}. The
remaining 94 percent give stable tori enclosing stable periodic orbits. The simultaneous existence of
one-dimensional stable orbits with two-dimensional tori and unstable three-dimensional orbits, comprising the
chaotic sea, all in close proximity to one another, was a surprise and heralded even more discoveries.

Soon after, in 2015, Wang and Yang showed that a stiffer higher-frequency thermostat, with $\dot \zeta
= 10(p^2 - 1)$, provides {\it knots}, not just simple overhand or trefoil knots, but a wide and complex
variety\cite{b68,b69}. In preparing this review we wondered whether or not such
structures could also be found in the prototypical $\alpha = 1$ case. Investigation revealed a
surprising complexity.  We soon found the stable periodic orbit of {\bf figure 9} with period
45.14 using $\alpha = 1$ . The initial values $(q,p,\zeta)$ were $(2.21,0,0)$. In principle, such
an orbit, embedded in three-dimensional space, can be analyzed according to a classical field of
mathematics, ``knot theory''.

Three-dimensional knots are now conventionally classified according to the number of crossings found
in their two-dimensional projections\cite{b41}. Wikipedia states that there are over a million topologically
distinct knots with 16 crossings, so that the knot shown in {\bf figure 9}, with its 36 crossings, is
likely well beyond the patience of even the most earnest of investigators. But this knot has a
simplifying two-fold symmetry shown at the left. The initial half-period, $0 < t < 22.57$, is a mirror
image of the second, $22.57 < t < 45.14$.

The two simpler half-period knots show only nine crossings. And manipulation of a plastic-chain model of
them soon results in a topologically equivalent half-knot with just eight crossings, well within the
range of analytic work as there are only 21 topologically different eight-crossings knots. Because knot
theory is an absorbing, challenging, and well-developed branch of mathematics, we expect it to contribute
to the statistical mechanics of nonequilibrium systems through the study of models like the thermostatted
oscillator considered here.

Let us now take leave of these singular periodic orbits to consider the fascinating world of ergodic
Lyapunov-unstable oscillator dynamics.  Ergodicity implies exploring the entire stationary distribution,
for almost all initial conditions. Such dynamics can and do in fact achieve Nos\'e's original
goal of reproducing the entire canonical ensemble. The simplest of such dynamics was discovered by
Clint Sprott in 2018\cite{b63}. Its less-simple precursors date back to the 1990s.

\subsection{Double Cross Sections / Numerical Methods}

An even simpler route to simulating stationary nonequilibrium systems is to make the imposed
temperature a function of coordinate. This can be accomplished with just the single degree of
freedom describing a one-dimensional oscillator. To illustrate this idea consider a smooth
temperature profile reminiscent of a shockwave:
$$
T(q) = 1 + \epsilon \tanh (q) .
$$
The oscillator friction coefficient $\zeta$ can then react to the local temperature $T(q)$ :
$$
\{ \ \dot q = p \ ; \ \dot p = -q - \zeta p \ ; \ \dot \zeta = p^2 - T(q) \ \} \ .
$$
In the following subsection we will consider a surprise which results from this simple nonequilibrium
problem, an astonishing complexity of oscillator orbits in the form of topological {\it knots},
an active mathematical field, with 1898 arXiv papers now listed under ``knot theory''.

Two of these occurred in 1997\cite{b57} and 2014\cite{b62}.  With Harald Posch, in 1997, Bill found that
exposing a harmonic oscillator to a coordinate-dependent temperature, $T(q) = 1 + \epsilon \tanh(q)$,
can result in another qualitatively different solution type, fractal (fractional-dimensional) phase-space
distributions.  Such fractals are ``dissipative'', with time-averaged vanishing phase-space volume in
response to overall positive ``friction'' from the control variable $\zeta$ :
$$
\langle \ \dot \otimes/\otimes \ \rangle = \langle\ -\zeta \ \rangle < 0 \longleftrightarrow
\langle\ p^3/2 \ \rangle \propto -\nabla T < 0 \ [ \ {\rm Fractal \ Solutions} \ ] \ .
$$
These chaotic trajectories invariably converge to stationary nonequilibrium flows of kinetic energy
$p\times(p^2/2)$ in the hot-to-cold direction. Though Lyapunov unstable, with chaos, the flows are
stable from the thermodynamic viewpoint, satisfying the Second Law.

With Clint Sprott, in 2014, we discovered that a pair of dissimilar two-dimensional conservative tori can
coexist, while simultaneously interlocked with a nonequilibrium dissipative limit cycle. On the one hand,
like the phase-space fractals, the limit cycle supports dissipation, phase-volume loss, and net hot-to-cold
heat flow.  On the other hand, the mutually interlocked conservative tori have no net heat flow,
$\langle \ p\times(p^2/2) \ \rangle$ -- nor do they show any tendency for phase-volume loss:
$$                                                                                                                                                                                                   
\langle \ -\dot\otimes/\otimes \ \rangle=-\langle \ (\partial \dot p/\partial p) \ \rangle                                                                                                           
= +\langle \ \zeta \ \rangle \equiv 0 \ [ \ {\rm Conservative \ Tori} \ ] \ .                                                                                                                        
$$

\begin{figure}
\includegraphics[width=2.5 in,angle=+90.]{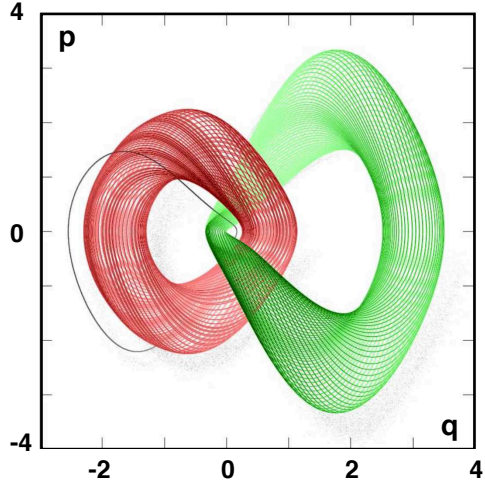}
\caption{
$\{ \ \dot q = p \ ; \ \dot p = -q - \zeta p \ ; \ \dot \zeta = p^2 - [ \ 1 + 0.42\tanh(q) \ ] \ \}$ with
initial conditions $(-2.7,0,0)$  [limit cycle], $(-2.3,0,0)$, and $(+3.5,0,0)$ [ tori ] showing the interlocked
nature of the three distinct solutions of exactly the same nonequilibrium Nos\'e-Hoover motion equations.
}
\end{figure}

{\bf figure 11} shows this situation where the oscillator dynamics of the friction coefficient $\zeta$ is
constrained by a moderate imposed local temperature, with $\epsilon = 0.42$:
$$                                                                                                                                                                                                   
T(q)= 1 + \epsilon \tanh(q) \rightarrow \nabla T = \epsilon/ \cosh^{2}(q) :                                                                                                                          
$$
$$
\{\ \ \dot q = p \ ; \ \dot p = - q - \zeta p \ ; \ \dot \zeta = p^2 - T(q) \ \} \
[ \ {\rm Nonequilibrium \ Oscillator} \ ] \ .
$$
Runge-Kutta integration of these motion equations with $\epsilon = 0.42$ gives numerical estimates for
the time-averaged loss of phase volume and the heat flux:
$$                                                                                                                                                                                                   
\langle \ \dot\otimes/\otimes \ \rangle = \langle \ (\partial \dot p/\partial p) \ \rangle =                                                                                                         
\langle \ -\zeta \ \rangle = -0.105 \ ; \ \langle \ p^3/2 \ \rangle = -0.259 \ .                                                                                                                     
$$
Notice again that each of the three solutions pictured is interlocked with the other two.

Graphical explorations in three or four dimensions can be based on sections of or projections onto
two-dimensional surfaces. In the absence of dissipation the Hoover-Holian and Martyna-Klein-Tuckerman
oscillators equations have the same stationary solution, Gaussian in all four variables $(q,p,\zeta,\xi)$.
The presence of an imposed temperature gradient can cause the Gaussians to condense onto fractal attractors
of reduced information dimension.

Let us illustrate with a moderate temperature gradient, $T(q) = 1 + 0.25\tanh(q)$.  We incorporate this
feature into the equations of motion as follows:
$$                                                                                                                                                                                                   
\{ \ \dot q = p \ ; \ \dot p = -q - \zeta p - \xi p^3 \ ; \ \dot \zeta = p^2 - T(q) \ ; \                                                                                                                 
\dot \xi = p^4 - 3p^2 \ \} \ [ \ {\rm HH} \ ] \ ;                                                                                                                                                    
$$
$$                                                                                                                                                                                                   
\{ \ \dot q = p \ ; \ \dot p = -q - \zeta p \ ; \ \dot \zeta = p^2 - T(q) - \xi \zeta \ ; \                                                                                                               
\dot \xi = \zeta^2 - T(q) \ \} \ [ \ {\rm MKT} \ ] \ .                                                                                                                                                  
$$

The structure of the resulting fractals can then be explored through double cross sections in which two of
the four time-dependent variables take on specified values as is illustrated in {\bf figure 12}.

\begin{figure}
\includegraphics[width= 2 in,angle=+90.]{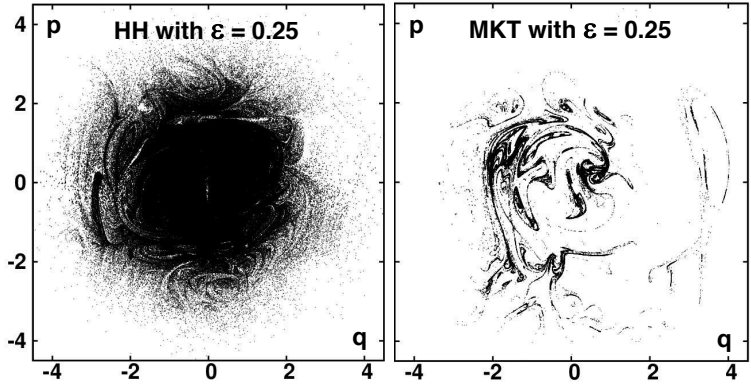}
\caption{
Double $p(q)$ conducting oscillator cross sections with $\zeta^2 + \xi^2 < 0.00001$. The
greatly enhanced equilibration of the stiffer Hoover-Holian thermostat relative to the gentler
Martyna-Klein-Tuckerman thermostat is apparent.
}
\end{figure}

\subsection{MKT, HH, 0532, and Sprott Routes to Ergodicity}

Because Nos\'e's goal, a deterministic dynamics reproducing Gibbs' canonical distribution, was eluded
by the prototypical oscillator problem it was natural to modify Nos\'e-Hoover mechanics in an attempt
to achieve ergodicity. Two successful approaches\cite{b28,b48} added a second control variable to $\zeta$:
$$
\{ \ \dot q = p \ ; \ \dot p = -q - \zeta p  \ ;
\ \dot \zeta = p^2 - 1 - \xi \zeta \ ; \ \dot \xi = \zeta^2 - 1 \ [ \ {\rm MKT} \ ] \ ;        
$$
$$
\{ \ \dot q = p \ ; \ \dot p = -q - \zeta p - \xi p^3 \ ;
\ \dot \zeta = p^2 - 1 \ ; \ \dot \xi = p^4 -3p^2 \ [ \ {\rm HH} \ ] \ .        
$$
Both approaches trace out exactly the same stationary state, a Gaussian in all four variables.
The Martyna-Klein-Tuckerman approach can be generalized to add ``chains'' of thermostat variables.
The Hoover-Holian approach can also be generalized further, but only a little, and at the expense
of excessive stiffness, by adding control of the sixth moment, $\langle \ p^6 \ \rangle$.

It is interesting to note that the fluctuations in the rate at which phase space is explored,
despite the identical distributions, are much larger in the HH case.  The mean-squared value
$\langle \ \dot\xi^2 \ \rangle$ is $\langle \ p^8 - 6p^6 + 9p^4 \ \rangle = 105 - 90 + 27 = 42$
for the HH equations and only $3 -2 + 1 = 2$ in the MKT case.  The relative stiffness of the
Hoover-Holian equations is the underlying reason why the sixth-moment analog of the HH equations
was long thought to be unstable.

The search for ergodic thermostats based on a single control variable rather than two, eventually
bore fruit.  Brute-force exploration of parameter space revealed that the ``0532'' oscillator, with
``weak control'' of both the second and fourth moments, is in fact ergodic:
$$
\{ \ \dot q = p \ ; \ \dot p = -q - \zeta[ \ 0.05p + 0.32p^3 \ ] \ ;
\ \dot \zeta = 0.05(p^2 - 1) + 0.32(p^4 -3p^2) \ \} \ [ \ 0532 \ ] \ .
$$  
In 2017 another single-control approach was the result of Tapias-Bravetti-Sanders' prize-winning variant
of the Nos\'e-Hoover oscillator\cite{b65} :
$$
\{ \ \dot q = p \ ; \ \dot p = -q - 10 \tanh(5\zeta)p \ ; \ \dot \zeta = p^2 - 1 \ \} \ 
[ \ {\rm TBS} \ ] \ .
$$
The next year Sprott, noting that a limiting case of TBS control is ``bang-bang'' (or ``on-off'')
control, suggested switching the friction coefficient discontinuously, $\pm\alpha \longleftrightarrow
\mp\alpha$, based on the kinetic temperature's history\cite{b63}. Sprott termed
this singular model a``Signum Thermostat'' as it depends only on the sign and not the magnitude of
the friction coefficient $\zeta$ :
$$
\{ \ \dot q = p\ ; \ \dot p = -q - \alpha ( \ \zeta/|\zeta| \ )p \ ; \ \dot \zeta = p^2 - 1 \ \} \ ,
[ \ {\rm Sprott's \ Signum \ Oscillator} \ ] \ .
$$
Sprott found that a sufficiently large value of $\alpha > 1.7$ provides an ergodic oscillator.  This problem
is instructive in that programming a precise interpolation to find the times at which $\zeta$ vanishes
just adds to the difficulty of carrying out an adaptive integration.  It is simpler and straightforward
to solve the motion equation $\dot p = -q - \alpha \tanh(400\zeta)p$ rather than the singular signum limit.
We have adopted this simplification in {\bf figure 13}.

\begin{figure}
\includegraphics[width=1.5 in,angle=+90.]{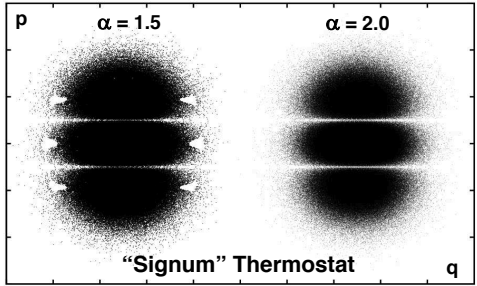}
\caption{
$\{ \ \dot q = p \ ; \ \dot p = -q -\alpha (\zeta/|\zeta|)p \ ; \ \dot \zeta = p^2 - 1 \ \}$. A
hyperbolic tangent approximation to Sprott's ``Signum'' thermostat applied to the harmonic oscillator:
$(\zeta/|\zeta|) \rightarrow \tanh(400\zeta)$. These $(q,p)$ Poincar\'e sections with $\zeta = 0$
were generated with one billion timesteps, $dt = 0.001$. The six holes to the left correspond to
stable periodic tori.  With $\alpha = 2$ the motion is ergodic.
}
\end{figure}

In none of these cases is ergodicity proven.  But several convincing numerical tests can be applied.
In 2015 we applied six of them to the MKT oscillator\cite{b35}. TBS provided more. Three of these tests are
relatively simple: [1] A Gaussian distribution with unit standard deviation should reproduce its
nonzero even moments, $\{ \ 1, \ 3, \ 15, \ 105, \ \dots \ \}$; [2] A Poincar\'e section cutting
through the distribution must reveal no holes; and [3] The largest Lyapunov exponent $\lambda_1$
must have the same long-time-averaged positive value independent of the initial condition. All of
these computational tests are relatively simple to administer.

A relatively stringent test of the four-dimensional Gaussians' ``no holes'' criterion can be based
on the construction of a ``double cross section'', plotting values of two of the four variables
whenever the remaining two are sufficiently close to zero. Points for the $(q,p)$ section of the MKT
oscillator can be collected whenever $(\zeta^2+\xi^2) < 0.00001$, for instance. A convincing
Gaussian results.  On the other hand points for the $(\zeta,\xi)$ section with $(q^2+p^2) < 0.00001$
look rather peculiar\cite{b52}. This is because there is no motion perpendicular to the $(q,p)$ plane
where both variables vanish, so that
$$
\{ \ \dot q = 0 \ ; \ \dot p = 0 \ ; \ \dot \zeta = -1 - \zeta\xi \ ; \ \dot \xi = \zeta^2 - 1 \ \} \ .
$$
These $(\zeta,\xi)$ equations are isomorphic to those of a falling particle, with speed $\zeta$,
controlled by a friction coefficient $\xi$ maintaining a ``kinetic temperature''
$\langle \ \zeta^2 \ \rangle$ of unity.  Let us describe that closely-related problem next.

\subsection{Nos\'e-Hoover Mechanics for the Falling Particle}

Our book written in Japan in 1989-1990, {\it Computational Statistical Mechanics}\cite{b25}, contains
an interesting example of Nos\'e-Hoover mechanics in which coordinates make no explicit appearance.
Coordinates can instead be introduced implicitly by integrating the momentum, $p = \dot q$.  Their
absence is reminiscent of the disappearance of Nos\'e's time-scaling variable $s$, which can be
calculated from the Nos\'e-Hoover equations by integrating the friction coefficient, $\zeta =
(\dot s/s)$. Now to the Nos\'e-Hoover example.

We consider a thermostatted particle with momentum $p$ under the influence of a constant external
gravitational field and a time-reversible frictional force $-\zeta p$. Just as in the oscillator
problems we simplify this example by choosing the mass, gravitational field strength, thermostat
frequency, temperature, and the geometric dimensionality of the problem  all equal to unity. The
prototypical equations of motion for the falling particle become
$$
\{ \ \dot p = -1 -\zeta p \ ; \ \dot \zeta = p^2 -1 \ \} \ [ \ {\rm Falling \ Particle} \ ] \ .
$$
Compare these to the motion equations in the $\{ \ q=0,p=0 \ \}$ plane for the MKT oscillator:
$$
\{ \ \dot \zeta =  - 1 - \xi \zeta \ ; \ \dot \xi = \zeta^2 - 1 \ \} \
[ \ {\rm MKT \ Oscillator \ in \ the} \ (0,0,\zeta,\xi) \ {\rm plane} \ ] \ .
$$
The two problems, a falling particle and an ergodic oscillator's $( \ 0,0,\zeta,\xi \ )$ plane motion
are one and the same! All these equations are time-reversible, in that any solution with positive
$dt$ can be used to find a solution with negative $dt$ by changing the signs of $p$, $\zeta$, and
$\xi$.

\begin{figure}
\includegraphics[width=2.5 in,angle=+90.]{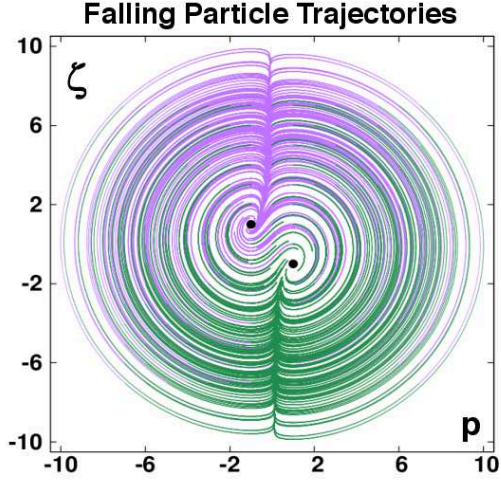}
\caption{
Time histories of 225 points initially located on a square grid from $-7$ to $+7$ in $p$ and $\zeta$,
followed for a time of $\pm3 = 3000dt$ forward (purple) and backward (green) toward the attractor at
$(-1,+1)$ and the repellor at $(+1,-1)$, both indicated by black dots.
}
\end{figure}

Both sets of motion equations have two fixed points, the unstable repellor ``source'', at $(p,\zeta)$
or $(\zeta,\xi) = (+1,-1)$ and the stable attractor ``sink'' at $(p,\zeta)$ or $(\zeta,\xi) = (-1,+1)$.
The repellor and attractor for the dissipative falling-particle flow are just ``fixed points'' in this
simplest example problem rather than the complex fractal objects familiar from more complicated flows.
The flow equations describe a time-reversible dissipative repellor-to-attractor flow from $(+1,-1)$
to $(-1,+1)$. See {\bf figure 14}. Changing the signs of both dependent variables as well as the time
generates trajectories approaching the attractor (purple) and the repellor (green).

A linear stability analysis of the flow in the neighborhood of the fixed points provides a detailed
understanding of the problem.  We begin at the attractor $(-1,+1)$ and describe the offset from that fixed
point by introducing two infinitesimal offset variables,
$(\delta_p,\delta_\zeta)$: $\delta_p \equiv p + 1$ and $\delta_\zeta \equiv \zeta - 1$. From the
Falling Particle equations above we then find the equations of motion for $\delta_p$ and $\delta_\zeta$ :
$$
\dot \delta_p \simeq -\zeta \delta _p - p\delta _\zeta \simeq - \delta_p + \delta _\zeta \longrightarrow 
\ddot \delta_p = -\dot \delta_p - 2\delta_p \ ;
$$
$$
\dot \delta_\zeta = 2p \delta_p \simeq - 2 \delta_p \longrightarrow
\ddot \delta_\zeta \simeq - 2\dot \delta_p \simeq 2\delta_p - 2\delta_\zeta
= - \dot \delta_\zeta - 2 \delta_\zeta \ . 
$$
The two linear second-order equations for $\delta_p$ and $\delta_\zeta$ have solutions proportional
to $e^{i\omega t}$ where substitution of that form of the solution gives the result :
$$
{\textstyle 
\omega^2 - i\omega - 2 = 0 \longrightarrow \omega = (i/2) \pm \sqrt{7/4} \ .
}
$$
{\bf figure 15} displays the logarithms of the {\it squared} offset components (so as to have real
logarithms) for both the repellor and the attractor.  Because for small offsets the dependence is
proportional to $e^{\pm t/2}$ the logarithms of the squares have a slope of unity when plotted as
functions of time. Counting the oscillations confirms the accuracy of the linear analysis just given.

\begin{figure}
\includegraphics[width=2.5 in,angle=+90.]{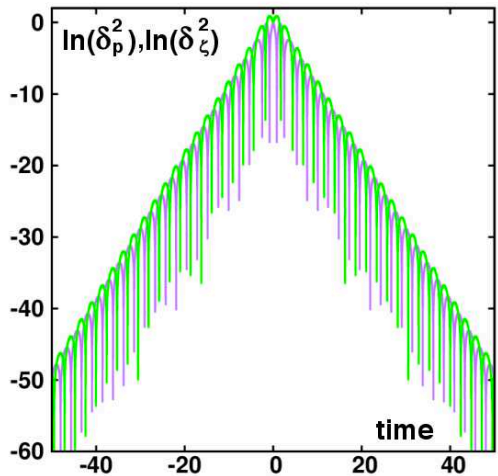}
\caption{
Logarithms of the squared offsets $\delta_p^2$ and $\delta_\zeta^2$ from the repellor and attractor
starting at (0,0) and going both forward to the attractor and backward toward the repellor for a
time of 50.  The slope of unity and the oscillation frequency from the 21 maxima corresponding to
10.5 periods agree perfectly with the linear stability analysis given in the text.
}
\end{figure}

The falling-particle flow itself has an interesting and stimulating stationary global solution for
the probability density $f(p,\zeta,t)$ which can be written in either of two forms. Here are both:
$$
\textstyle{
f(p,\zeta,t) \propto e^{ -(p^2/2) - (\zeta^2/2) - \int_0^t p(t^\prime) dt^\prime } \equiv
 e^{ + \int_0^t \zeta (t^\prime) dt^\prime } \ .
}
$$
We can confirm the two expressions simultaneously by showing that there is a constant of the motion,
$C$ for the flow:
$$
\textstyle{
{\cal C} \equiv (p^2/2) + (\zeta^2/2) +
\int_0^t p(t^\prime) dt^\prime + \int_0^t \zeta (t^\prime) dt^\prime \longrightarrow
}
$$
$$
\textstyle{
\dot {\cal C} = p\dot p + \zeta \dot \zeta + p + \zeta =
p(-1 - \zeta p) + \zeta(p^2 - 1) + p + \zeta \equiv 0 \ .
}
$$
{\bf figure 15} describes a particular solution of these model equations centered on (0,0) and
proceeding either forward or backward by choosing the appropriate sign for the timestep
$dt$.

By introducing more complexity in the falling-particle model, with a more elaborate
thermostat (HH or MKT) and with two space dimensions rather than just one, a variety
of interesting nonequilibrium states can be constructed and analyzed. This appears to
be an excellent problem for elaboration and exploration. We recommend it for further study.

\section{Lyapunov Exponents, Heat Flow, Fractals, and Shockwaves}

Linear differential equations have exponential solutions, real, imaginary, or complex, as typified by
biological growth, mechanical dissipation, or periodic oscillations. Driven, damped, and underdamped
oscillators provide the simplest examples of these three possibilities. Manybody classical systems are
typically ``Lyapunov unstable''. This means that a small change is amplified exponentially fast by the
dynamics. In an $N$-dimensional phase-space representation of the dynamics we can imagine defining a
spectrum of growth rates $\lambda_1 \dots \lambda_N$ by measuring the expansion or contraction of a
comoving hypersphere or hypercube. Practical measurement algorithms for these exponents were developed
by Shimada and Nagashima\cite{b61} and Benettin's group\cite{b7} forty years ago, and are relatively
easy to implement, requiring only continuous attention to maintain the orthogonality of the vectors
representing the directions in which the dynamical expansions and contractions occur.  Let us consider
here a familiar three-dimensional example, the thermostatted Nos\'e-Hoover oscillator.  This problem
has an unfamiliar moral.

In Hamiltonian mechanics the conservation of phase volume, derived through Liouville's Theorem,
$(\dot \otimes/\otimes) = \sum \lambda_i\equiv 0$, guarantees that the sum of the all the Lyapunov exponents
vanishes for Hamiltonian systems.  Time-reversibility correctly suggests that these exponents come in
pairs, $\{ \ \pm \lambda \ \}$, with the signs expressing expansion or contraction corresponding to
following the motion either forward or backward in time. Although the time-reversibility of thermostatted
nonequilibrium systems certainly suggests this same pairing of positive and negative exponents, in
practice the possibility of dissipation, forming a chaotic ``strange attractor'' in the state space, is
seized upon by the dynamics. The Nos\'e-Hoover oscillator in a temperature gradient provides a simple
example problem where the ``time-reversible'' ``attractor'' actually gives way to an irreversible
one-dimensional limit cycle\cite{b33}.
\begin{figure}
\includegraphics[width=2.5 in,angle=+90.]{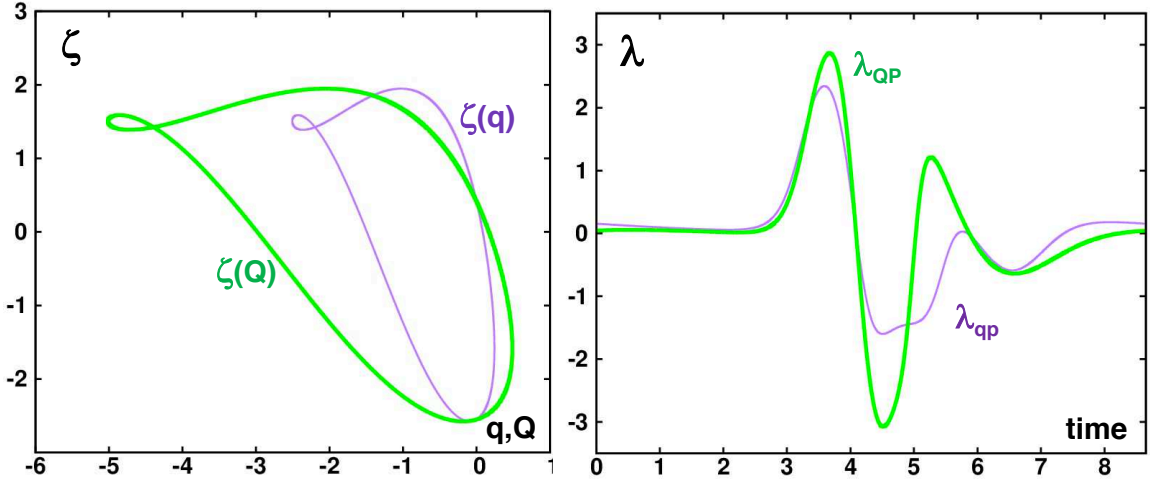}
\caption{
This figure demonstrates that the local Lyapunov exponent $\lambda_1(t)$ depends on the coordinate system
choice for the Nos\'e-Hoover oscillator, corresponding to choosing $s=$ 1 or 2.
}
\end{figure}
In {\bf figure 16} two different coordinate systems describe the same periodic orbit:
$$
\{ \ \dot q = p \ ; \ \dot p = -q - \zeta p \ ; \ \dot \zeta = p^2 - T(q) \ \} \ ; \
T = 1 + 0.5\tanh(q) \ {\rm with} \ s=1 \ ;
$$
$$
\{ \ \dot Q = 4P \ ; \ \dot P = -(Q/4) - \zeta P \ ; \ \dot \zeta = 4P^2 - T(Q) \ \} \ ; \
T = 1 + 0.5\tanh(Q/2) \  {\rm for} \ s=2 \ .
$$
The relationship between the two coordinate systems illustrated here is the simplest
possible: $(q,p) \rightarrow (Q/s,sP)$. Here both $s=1$ (purple) and $s=2$ (green) are illustrated.
The two limit cycles, both with period 8.65, were obtained after discarding ten million timesteps
$dt = 0.001$ from the initial conditions $(q,p,\zeta)=(0,5,0)$. The subsequent time dependence of
the friction coefficient $\zeta(t)$ (at the left) is independent of $s$ but the local values of the
time-dependent exponents depend upon the scale factor $s$, with the largest local Lyapunov exponent
illustrating this dependence at the right.

Despite the time-reversibility of these motion equations symmetry breaking promptly finds the limit-cycle
solution illustrated in {\bf figure 16}. Investigation reveals that the largest of the three Lyapunov
exponents has time-averaged value zero, and can be thought of as reflecting the time-averaged lack of
expansion or contraction of a vector parallel to the periodic dynamical $(q,p,\zeta)$ trajectory.

A thorough dynamical description of the motion is more complicated than the single-exponent example of
{\bf figure 16}.  Computing all three Lyapunov exponents requires the solution of 12 ordinary differential
equations. Three of the motion equations describe an underlying reference trajectory. The additional nine
characterize the directions of three comoving and corotating orthogonal offset vectors. The offsets are
``small'' displacements from the reference trajectory. The vectors, typically of length $10^{-5}$ when
the reference trajectory variables are of order unity, can be described in the same three-dimensional
$(q,p,\zeta)$ state space as the reference trajectory. Alternatively the equations of motion can be
linearized around the values of the nearby reference trajectory:
$$
\{ \ \dot \delta_q = \delta_p \ ; \ \dot \delta_p = -\delta_q -\zeta\delta_p - p\delta_\zeta \ ; \
\dot \delta_\zeta = 2p\delta_p - (dT/dq)\delta_q \ \} \ ,
$$
with infinitesimal offset vectors taken to be of unit length\cite{b58}.

In either case constraints maintaining the lengths and orthogonality of the three vectors can be imposed
on the motion by six Lagrange multipliers\cite{b27}, three maintaining orthogonality and three maintaining
the vector lengths, with the latter giving the three local Lyapunov exponents, $\{ \ \lambda_i(t) \ \}$.
When time-averaged these exponents, $\{ \ 0, \ -0.0715, \ -0.2225 \ \}$, sum to give the time-rate-of-change
of the comoving three-dimensional volume, showing that an infinitesimal cylinder moving with the trajectory
shrinks radially toward the limit cycle ;
$$
\langle \ (\dot \otimes/\otimes) \ \rangle = \langle \ (\partial \dot q/\partial q)+(\partial \dot p/\partial p)+
(\partial \dot \zeta/\partial \zeta) \ \rangle = \langle \ 0 - \zeta + 0 \ \rangle =
\lambda_1 + \lambda_2 + \lambda_3 = -0.2940 \ .
$$

The straightforward ``Gram-Schmidt'' algorithm maintains the orthogonality and orthonormality of the
offset vectors and can be applied at every timestep.  The first offset vector, which determines the
exponent $\lambda_1(t)$, is simply rescaled in length at every timestep so as to determine the local
exponent:
$$
\lambda_1 \equiv (1/dt)\ln[ \ (\delta_{\rm before}/\delta_{\rm after}) \ ] \ .
$$
The second offset vector then undergoes two constraints: [1] orthogonality with $\delta_1$ is first
imposed, by subtracting that fraction of $\delta_2$ which parallels $\delta_1$.  [2] the length of
$\delta_2$ is then rescaled, giving the second local Lyapunov exponent $\lambda_2(t)$:
$\lambda_2 \equiv (1/dt)\ln[ \ (\delta_{\rm before}/\delta_{\rm after}) \ ] \ .$
The alternative Lagrange-multiplier constraints require only occasional applications of the Gram-Schmidt
orthonormality algorithm to offset the inevitable effects of roundoff error.  In manybody systems the
Lyapunov spectra often exhibit a powerlaw form reminiscent of the Debye Model's vibrational
frequencies of a continuum\cite{b56}. The time-averaged spectrum of Lyapunov exponents, unlike the
local instantaneous spectra, has a very interesting connection to the Second Law of Thermodynamics.  We
describe it in the next subsection for a family of one- and  two-dimensional manybody problems exhibiting
heat conductivity and interesting Lyapunov spectra.

\subsection{Fractal Dimensionality Loss for 24 $\phi^4$ Particles\cite{b30}}

The thermostat forces generalizing Nos\'e's work to nonequilibrium systems make it possible to study
``mesoscopic'' systems, small enough for an atomistic description but large enough to be described with
continuum concepts.  The simplest such system is the $\phi^4$ model, a regular lattice, one-, two-,
or three-dimensional, of particles tethered to lattice sites with a quartic potential and with
additional nearest-neighbor Hooke's-Law harmonic forces. In 2004 two-dimensional $\phi^4$ models
provided an excellent testbed for comparing a variety of thermostat forces. We carried out a thorough
comparison of seven distinct thermostat types, all of them applied to two-dimensional models with four
thermostatted cold particles at a cold-end kinetic temperature of 0.5 and four thermostatted hot particles
at a hot-end temperature of 1.5. 16, 32, or 64 Newtonian particles were sandwiched between the two
thermostatted columns\cite{b31}. Series of simulations for the different system sizes and different
thermostats suggested that the various approaches would agree for large systems.  Deviations from this
imagined large-system limit could be estimated for each of the thermostats.  In the end this work reached
a simple conclusion :
\begin{quote}
``The simplicity of thermostats based on the second moment of the velocity distribution
and simply connected to irreversible thermodynamics recommends the use of Nos\'e-Hoover thermostats
whenever possible.''
\end{quote}

\subsection{An Educational One-Dimensional Model of Heat Flow}

The encouraging nature of the two-dimensional simulation results led us to subsequent investigations of
somewhat longer one-dimensional $\phi^4$ chains\cite{b36}. {\bf figure 17} describes the Lyapunov
instability of a typical 24-atom chain.  The figure shows the largest sixteen Lyapunov exponents from
the spectrum of 50. The 50-dimensional phase space is composed of 24 coordinates, 24 momenta, and
two Nos\'e-Hoover friction coefficients.  The two thermostats maintain time-averaged kinetic
temperatures of 0.003 and 0.027 at the two ends of the 24-particle chain. The tethering force constant,
the Hooke's-Law force constant, and the particle masses were all taken equal to unity.  To illustrate,
the motion equations of the first (cold) particle and its neighbor in the chain are as follows:
$$
\dot \zeta_1 = \dot q_1^2 - 0.003 \ ; \ \ddot q_1 = -q_1^3 + (q_2-q_1) - \zeta_1 \dot q_1 \ ; \
\ddot q_2 = -q_2^3 + (q_1 + q_3 - 2q_2) \ . 
$$
Here the particle coordinates $\{ \ q \ \}$ are measured relative to their fixed lattice positions. The
Lyapunov exponents summed-up at the right in {\bf figure 17} show that a 15-dimensional hypervolume in
the phase space expands exponentially with time while a 16-dimensional hypervolume contracts.  Evidently
the fractal dimension of the strange attractor (between 15 and 16) lies below that of the phase space
by a ``dimensionality loss'' of at least 34, about 70 percent of the total. This feature of
time-reversible dynamical systems is striking in its significance and generality. With this specific
microscopic case in mind let us consider a corresponding ``thought experiment'' close to macroscopic
thermodynamics and its Second Law.

\begin{figure}
\includegraphics[width=2.5 in,angle=-90.]{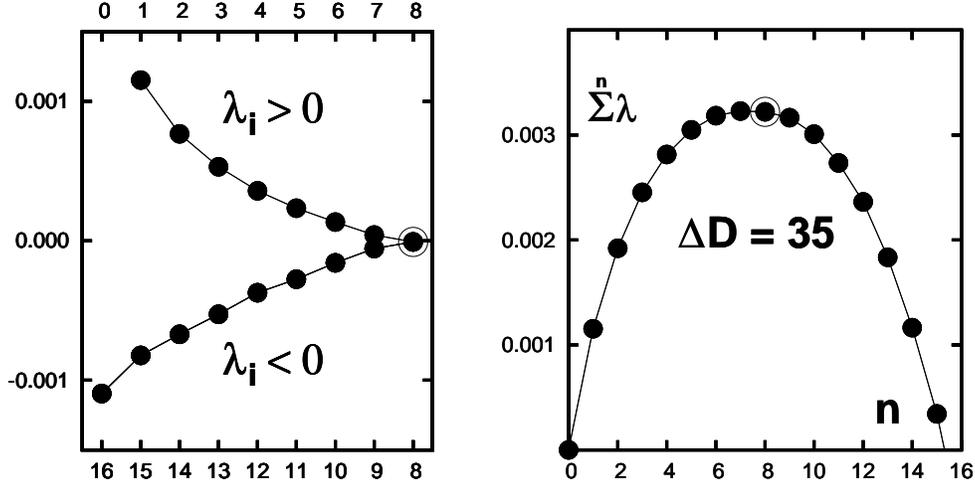}
\caption{
A part of the 50-exponent Lyapunov spectrum of a conducting $\phi^4$ chain
is shown at the left where Nos\'e-Hoover thermostats maintain boundary temperatures (of a single
cold and single hot particle) at 0.003 and 0.027.  At the left we see seven positive exponents, a
zero (corresponding to the lack of a time-averaged expansion or contraction along the trajectory
direction in phase space), and eight negative exponents, just enough to furnish a negative overall
sum, placing the dimensionality of the steady-state strange attractor between 15 and 16. At the right
the sums of exponents are plotted. For more details see references 30 and 36.
}
\end{figure}

\subsection{Macroscopic Heat Flow and the Second Law of Thermodynamics}

Imagine a conducting solid with many degrees of freedom connected to two time-reversible heat reservoirs
with different temperatures, one ``cold'' and the other ``hot'', just as in the 24-particle model
example above. Imagine that this macroscopic solid body obeys Fourier's Law, with a heat flux proportional
to the local temperature gradient, $Q \propto -\nabla T$.  We expect that on a time-averaged basis heat
would flow from the hot reservoir through the system to the cold reservoir at a rate proportional to the
temperature difference between the reservoirs. The consequent thermodynamic rate of increase in the
solid's entropy at the hot reservoir, $\simeq Q/T_{\rm hot}$, is then more than offset by the larger
rate of entropy decrease at the cold reservoir $\simeq Q/T_{\rm cold}$. The greater loss at
$T_{\rm cold}$ leads to the embarrassing conclusion that in a heat-conducting steady state the system
entropy inexorably diverges to minus infinity! The conventional cure for this straightforward conclusion
is to imagine an ``entropy production'' inside the system chosen exactly to offset the net entropy
loss at the reservoir-system boundaries.  Without this {\it ad hoc} assumption the entropy changes
{\it at} the boundaries would sum to zero rather than to the positive result required by the Second
Law: the entropy of the ``Universe'' (system plus reservoirs here) approaches a maximum.

Evidently the macroscopic analysis of this thought experiment is not particularly informative. If we
instead consider the alternative microscopic point of view, imagining that Nos\'e-Hoover heat reservoirs
maintain the hot and cold boundary temperatures, we can develop time-averaged equations relating the
energy extracted from the solid to the summed-up hot and cold friction coefficients and to the volume
change in phase space:
$$
\sum \langle \ \zeta p^2/mkT  \ \rangle \equiv \sum \langle \ \zeta  \ \rangle = \sum \langle \
(-\partial \dot p/\partial p) \ \rangle = \langle \ (-\dot \otimes/\otimes) \ \rangle
= \langle \ -(\dot S/k) \ \rangle \ .
$$
Here the time-averaged equality of $\zeta p^2$ and $\zeta mkT$ follows from the vanishing of the
averaged time derivative of a bounded quantity
$0=\langle \ \frac{d}{dt}(\zeta^2/2) = \zeta\dot \zeta \ \rangle \propto \langle \ \zeta(p^2-mkT) \ \rangle$.

The formation of fractal attractors and repellors in the phase space is thoroughly consistent with
macroscopic thermodynamics.  Nonequilibrium microstates, both the repellor and the attractor, are
both so rare as to have zero measure in the equilibrium phase space, due to dimensionality loss.
The fractal attractor is stable, with shrinking phase volume, while the repellor is not, making
its observation impossible. A simulation beginning with a reversed heat flow and reversed friction
coefficients quickly illustrates its numerical instability by returning to its mirror-image attractor.
The compatibility of the paired attractors and repellors in nonequilibrium systems became apparent in
1987\cite{b16,b26,b49} and supports the use of reversible thermostats in modelling nonequilibrium systems.
Simulations of shear and heat flow have a long history and a record of good agreement with
experimental values\cite{b6}. Let us turn to systems which are not yet so well understood, systems
in which the transport is nonlinear, shockwaves.

\subsection{Macroscopic and Microscopic Shockwave Models}

Nonequilibrium systems involve dissipation, irreversibility, and boundary conditions linking the system
to the outside world, which drives it away from equilibrium. In modelling nonequilibrium stationary states
the heat generated inside the system is extracted by external heat reservoirs. The simplest nonequilibrium
steady states, shear flow and heat flow, require transport of momentum and energy {\it through} such a
system. Such nonequilibrium steady states can be described by a variety of models, of which the macroscopic
Navier-Stokes equations and the microscopic Boltzmann Equation are the simplest representatives.  A host
of simulations of the viscosities shear and bulk, and the heat conductivity have been carried out and
found to give useful results despite the effects of boundaries.

Here we focus on phenomena for which the boundary conditions are purely equilibrium, as in the interface
between coexisting phases, but with the boundary between different states characterized by
different velocities.  Let us consider a steady ``one-dimensional'' (meaning planar) shockwave, with
cold low-pressure material transformed to a higher-pressure higher-temperature compressed state by motion
of a steady wave.  We begin by indicating a predictive path to the steady structure of such a wave.

\subsection{Macroscopic Shockwaves from the Navier-Stokes Equations\cite{b15,b22}}

\begin{figure}
\includegraphics[width=1.5in,angle=-90.]{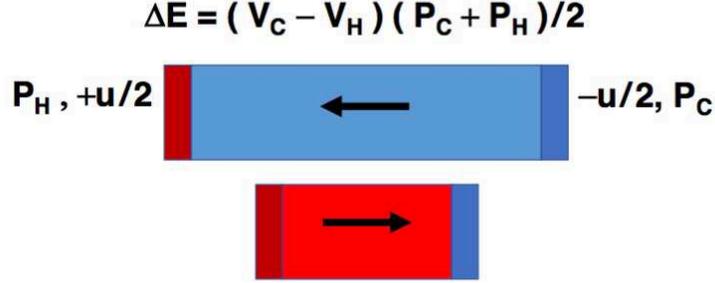}
\caption{
Illustrating the shockwave conversion of work to heat as described by the Rankine-Hugoniot
relation.  Left-moving cold fluid is transformed to right-moving hot fluid (with exactly the same
kinetic energy and with the additional ``internal'' energy $E_{\rm hot} - E_{\rm cold}$) through
the fluid's reaction to the work done on it by the two moving pistons:
$\Delta E = P_{\rm cold}\Delta V/2 + P_{\rm hot}\Delta V/2$.
}
\end{figure}

The two simplest transport processes appear together in sound waves, which decay, and in shockwaves,
where a steady input of fast-moving cold fluid is converted to a slower hotter output state by a localized
nonequilibrium process. The macroscopic Rankine-Hugoniot Equations, dating back to the nineteenth century,
are most easily derived by imagining the nonequilibrium conversion of a left-moving fluid to right-moving
with equal speeds but opposite velocities before and after.  As the kinetic energy is unchanged the
internal energy increase is equal to the work done:
$$
(P_{\rm hot} + P_{\rm cold})(1/2)(V_{\rm cold} - V_{\rm hot}) = (e_{\rm hot} - e_{\rm cold}) \ [ \
{\rm Rankine-Hugoniot \ Equation} \ ] \ .
$$

Though the Rankine-Hugoniot equation expressed in {\bf figure 18} correctly describes the energy changes
taking place in a steady
shockwave it provides no indication of the equilibration mechanism(s) responsible. Viewing the shockwave
compression process in a different coordinate frame, centered on the shockwave, can provide descriptions
well suited to comparison with simulations and shown in considerable microscopic detail in {\bf figure 19}.

In that figure cold fluid enters at the left and passes through the shockwave transition region to become a
denser, higher-pressure, higher-temperature ``shocked'' fluid, which exits at the right. In the
frame centered on the shockwave the flow is completely stationary, simplifying its description in terms of
simple models such as the microscopic Boltzmann Equation and the macroscopic Navier-Stokes equations. The
Navier-Stokes approach is the simpler of the two. A short computer program solving that model can be based on the three
macroscopic conservation laws, conservation of mass, momentum, and energy. We illustrate this next.

\begin{figure}
\includegraphics[width=2.5 in,angle=+90.]{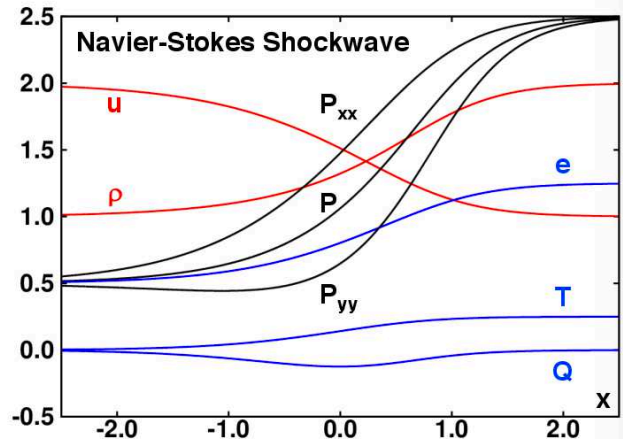}
\caption{
This simplest possible shockwave picture includes Newtonian viscosity and Fourier heat
conduction, as formalized in the Navier-Stokes equations.  The Runge-Kutta solution of the conservation
laws for stationary mass, momentum, and energy fluxes requires solving two ordinary differential
equations for the velocity and temperature profiles subject to equilibrium boundary conditions satisfying
the Hugoniot equation.  The equilibrium mechanical and thermal equations of state,
$P = \rho e = (\rho^2/2) +\rho T$,
together with Newtonian shear viscosity and Fourier heat conductivity, give a stationary shockwave
profile. We display the coordinate dependence of the mass density $\rho$ and fluid velocity $u$, the
pressure tensor $P_{xx},P_{yy}$ including the viscous contributions $\mp (du/dx)$, the energy per unit
mass $e$, the temperature $T$, and the heat flux $Q \equiv -(dT/dx)$.
}
\end{figure}

The mass flux $\rho u$ is necessarily constant throughout the steady state. This follows from the continuity
equation, $(\partial \rho/\partial t) = -\nabla \cdot (\rho u)$. The momentum flux, in addition to the comoving
component $P_{xx}$ parallel to the flow, includes the additional contribution $\rho u^2$ due to the net flow
velocity $u$ in the coordinate frame of {\bf figure 19}. A consequence of these two observations is that the
longitudinal pressure-tensor component {\it within} the wave varies linearly with volume, the ``Rayleigh Line'':
$$
P_{xx}(x) + (1/\rho)(\rho u)^2 = {\rm constant} \ [ \ {\rm Momentum \ Flux} \ ] \ ,
$$ 
as the mass flux $(\rho u)$ is constant in the wave. It is remarkable that the highly nonequilibrium
pressure-tensor component $P_{xx}$ can be determined by a relatively simple velocity measurement. 

A one-dimensional steady shockwave wave has stationary values of the mass, momentum, and energy fluxes
throughout the wave :
$$
\rho u, \ P_{xx} + \rho u^2, \ \rho u[ \ e + (P_{xx}/\rho) + (u^2/2) \ ] + Q_x \ .
$$
$P_{xx}$ and $Q_x$ are the longitudinal pressure and heat flux, $e$ is the internal energy per unit mass,
and u is the stream velocity, $u_s$ (for ``shock'') to the left of the stationary wave and $u_s - u_p$ (for
``shock minus particle'') to the right.  Although the equations appear complicated they are quite easy to solve
using the Runge-Kutta integrator. To do so integrate simultaneously the differential equations for $(du/dx)$ and
$(dT/dx)$.  We illustrate the integration here for a simple example problem.

In this Navier-Stokes model of a shockwave we adopt the following constitutive relations:
$$
P = \rho e = (\rho^2/2)+\rho T \ ; \ e= (\rho/2)+T \ ; \ (P_{xx} - P_{yy})/2 = -du/dx \ ; \ 
(P_{xx} + P_{yy})/2 = P .
$$
The mass, momentum, and energy flux have the values $(2, 4.5, 6)$.
For boundary conditions (cold to the left and hot to the right) we choose:
$$
u:[2\rightarrow 1] \ ; \ \rho:[1 \rightarrow 2] \ ; \ P : [1/2\rightarrow 5/2] \ ;
 \ e:[1/2\rightarrow 5/4] \ ; \ T :[0\rightarrow 1/4].
$$

A little numerical experimentation shows that the continuum equations for the velocity and temperature :
$$
\{ \ (du/dx) = P + \rho u^2 - 4.5 \ ; \ (dT/dx) = 2[ \ e + (P_{xx}/\rho) + (u^2/2) \ ] - 6 \ \} \ ,
$$
can be solved by starting at the ``hot'' side of the shockwave and integrating toward the ``cold'' side.
The logic of the program is as follows.  Given $u$ and $T$ compute $\rho$ from the mass flux, after which
the equilibrium pressure $P$ is known, which makes it possible to compute $(du/dx)$ from the momentum
flux. The internal energy is then known making it possible to compute the temperature gradient $(dT/dx)$
from the energy flux. This process provides the internal structure of the shockwave shown in {\bf figure 19}.
In the case shown in the figure the density doubles and the longitudinal pressure exceeds the transverse by
about a factor of three at the center of the wave.  Although twofold compression may seem extreme typical
shockwave experiments carried out in laboratory settings compress metals as much as fourfold at pressures
comparable to that at the center of the earth, a few million times normal atmospheric pressures.

\subsection{Remarks on Microscopic Shockwave Simulations}

Microscopic simulations have revealed two flaws in this Navier-Stokes model: [1] the nonequilibrium
dependence of energy on density and temperature in the shock necessarily deviates from the equilibrium value;
[2] the temperature itself within the shock is necessarily anisotropic. The kinetic temperature in atomistic
simulations of shockwaves\cite{b15} is easy to measure and is typically anisotropic, where the longitudinal
and transverse temperatures are proportional to $\langle \ p_x^2 \ \rangle \neq \langle \ p_y^2 \ \rangle$.
These two kinetic temperatures can be different within the shock by a factor of two or more in shockwaves of
moderate strength. A well-documented example is the 400 kilobar shockwave model of argon, twofold compressed
from the liquid\cite{b15}. The characterization of deviations from the linear-transport Navier-Stokes
assumptions is an active and rewarding research area, well suited to microscopic modelling.

By 1973 Ashurst had developed techniques for controlling boundary regions' velocities and kinetic temperatures
through time-reversible constraint forces. His viscosity and conductivity results agreed well with the somewhat
more cumbersome equilibrium Green-Kubo simulations\cite{b6,b47}. By 1980 several molecular dynamics simulations
of shockwaves, with as many as 4800 particles, had been carried out, making it possible to compare the
microscopic and macroscopic shockwave profiles\cite{b15,b44}.  This work was particularly welcomed by
those who interpretted laboratory measurements of shockwave and particle velocities to obtain equations
of state for real materials.  The findings from atomistic computer simulations that strong shockwave
thicknesses are only a few particle diameters, and later, that nonplanar sinusoidal-shaped shocks
promptly become planar\cite{b36}, confirm the accuracy of the assumptions underlying the Rankine-Hugoniot
equation and the equations of state obtained from it through laboratory velocity measurements.

\subsection{Three Routes to the Information Dimension for the N2 and N3 Maps}

Let us complete our view of interesting computational models by turning to deterministic ``maps'', the simplest
models of nonequilibrium systems. In maps discrete jumps in a model phase space, usually two-dimensional,
replace the continuous flows generated by ordinary differential equations.  Flows require at least three
dimensions plus nonlinearity for chaos.  With maps linearity and just two dimensions are enough. Chaos and fractal
structures can be achieved with simple linear mappings\cite{b29,b45} in a two-dimensional space.

Hopf introduced the Baker Map in 1937. Its action is reminiscent of dividing a clump of bread dough in half,
merging the pieces and then repeating the division and merging {\it ad infinitum}. Bill Vance pointed out to Bill
that a 45-degree rotation of that Baker Map serves to make it time-reversible. Adding twofold changes in area to
the 45-degree rotation to the Baker Map produces the leftmost N2 map of {\bf figure 20} and the resulting
excellent fractal shown at the left in {\bf figure 22} with its early two-iteration stage shown in {\bf figure 21}.
Maps like N2 are time-reversible and compressible, thus sharing the main features of chaotic dissipative flows
based on time-reversible versions or variations of classical mechanics.

\begin{figure}
\includegraphics[width=2 in,angle=-90.]{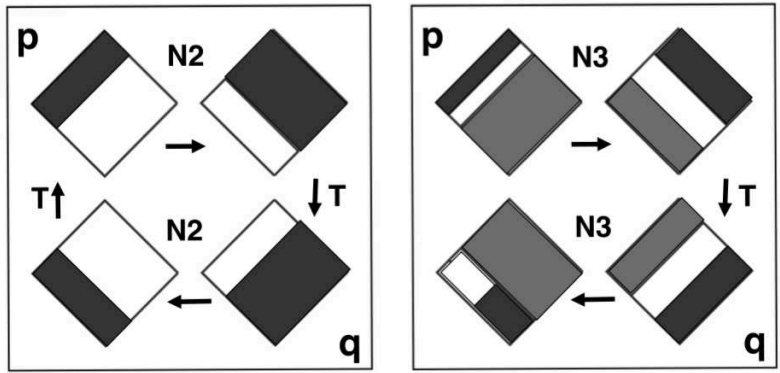}
\caption{
The time-reversible compressible Baker Map N2 and the irreversible Baker Map N3. Both maps provide the same
distributions of probability, in the fractal $q=p$ direction, after any finite number of iterations.
Nevertheless the fractal sets of points generated by these maps exhibit different information dimensions $D_I$.
}
\end{figure}

Time-reversibility of a map is analogous to the time-reversibility of Hamiltonian motion equations.  It
 implies that the inverse map N2$^{-1}$ is identical to carrying out a sequence of
three mappings, T*N2*T, where the time-reversal mapping $T(q,\pm p)\rightarrow(q,\mp p)$ just changes the sign of the
``momentum variable'' $p$. The reader can verify that the similar N3 map is not time-reversible.

The N2 and N3
mappings shown in {\bf figure 20} both map the southeastern two thirds of the $(q,p)$ domain into the
southwestern one third and the northwestern third into the northeastern two thirds. We began the study of N2
in 1998\cite{b38} and have recently considered it in more detail\cite{b39}. We found that three different
routes to the information dimension of the N2 fractal give at least three different results! To shed light on
this puzzle we developed and investigated the closely-related N3 map, for which all three routes agree.

\begin{figure}
\includegraphics[width=2 in,angle=-90.]{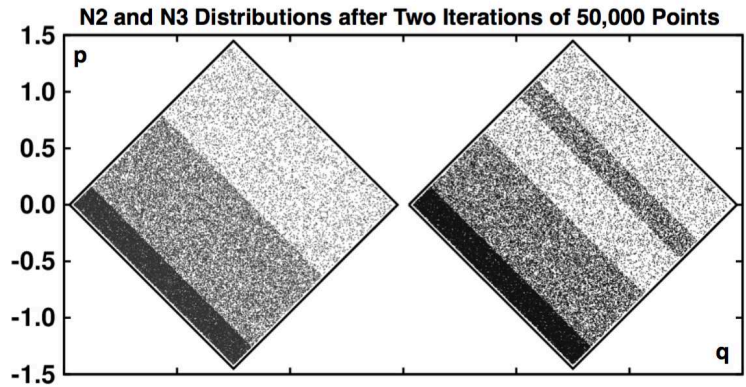}
\caption{
After two forward mappings of N2 and N3 the distributions of 50,000 random initial points occupy nine
strips of equal width parallel to the $q=p$ direction. The relative densities of the strips are 4 for
the densest strip, 1 for four of intermediate density, and 1/4 for the four least-dense strips. Both
have equal information dimensions, 1.7897 overall and 0.7897 in the fractal direction. See reference
38 for additional details.
}
\end{figure}

The N2 map has a straightforward linear form:
\begin{verbatim}                                                                                                      
if(q-p.lt.-4*d) qnew = + (11/ 6)*q - ( 7/ 6)*p + 14*d
if(q-p.lt.-4*d) pnew = - ( 7/ 6)*q + (11/ 6)*p - 10*d
if(q-p.ge.-4*d) qnew = + (11/12)*q - ( 7/12)*p -  7*d
if(q-p.ge.-4*d) pnew = - ( 7/12)*q + (11/12)*p -  1*d
[Reversible Nonequilibrium N2 Map use d = sqrt(1/72)]
\end{verbatim}
and its inverse mapping is no more complicated than the forward mapping:
\begin{verbatim}                                                                                                    
if(q+p.lt.-4*d) qnew = + (11/ 6)*q + ( 7/ 6)*p + 14*d
if(q+p.lt.-4*d) pnew = + ( 7/ 6)*q + (11/ 6)*p + 10*d
if(q+p.ge.-4*d) qnew = + (11/12)*q + ( 7/12)*p -  7*d
if(q+p.ge.-4*d) pnew = + ( 7/12)*q + (11/12)*p +  1*d
[Inversion of Reversible Nonequilibrium Baker Map N2]
\end{verbatim}
The details of the somewhat more complicated N3 map and its inverse can be found in reference 39.

In 1998 we adopted the conventional wisdom that the information dimension of the fractal generated by N2
was correctly given in terms of the maps' Lyapunov exponents by the Kaplan-Yorke conjecture\cite{b42},
$$
D_I \stackrel{?}{=}D_{\rm KY} = 1 - (\lambda_1/\lambda_2) \ [ \ {\rm Kaplan-Yorke} \ ] \ .
$$
Kaplan and Yorke interpolated between the growth rate of a one-dimensional length, $\lambda_1$, and the
decay rate of a two-dimensional area, $\lambda_1+\lambda_2$, to estimate the information dimension $D_I$
of a ``typical'' two-dimensional map. In the N2 case the stretching direction provides Lyapunov exponent
$\lambda_1$ and the shrinking fractal direction parallel to $q=p$ gives $\lambda_2$ :
$$
\lambda_1 = (2/3)\ln(3/2) + (1/3)\ln(3) = (1/3)\ln(27/4) \ ;
$$
$$
\lambda_2 = (2/3)\ln(1/3) + (1/3)\ln(2/3) = (1/3)\ln(2/27) \ [ \ {\rm N2 \ Exponents} \ ] \ .
$$
Together these give $D_{\rm KY} = 1 + 0.636514/0.867563 = 1.7337$ for the two-dimensional N2 map, with the
fractal direction's information dimension accounting for the 0.7337.

Another estimate for the information dimension comes from noting that each iteration of the map provides new
``information'' on a threefold smaller scale, with each region of constant probability ``prob'' giving a new
region with (2/3)prob and two regions with (1/6)prob each for a gain in information
$(2/3)\ln(3/2) + (1/3)\ln(6) = 0.270310 +0.597253$
corresponding to an information dimension of 0.7897 rather than 0.7337. A third and a fourth estimate come
from iterating a point, rather than regions, as many as a trillion times, and computing the dependence of
$\langle \ \ln({\rm prob}) \ \rangle$ on the bin size $\delta$.  Using inverse powers of 3 gives an estimate of
the fractal dimension 0.741 while inverse powers of 4 and 5 appear to be consistent with the Kaplan-Yorke
estimate 0.7337.  This inflationary wealth of estimates led us to consider the N3 map shown at the right in
{\bf figure 20}. {\bf figure 22} shows the fractals which result from the two mappings by iterating 100,000
times, starting from the central point $(q,p) = (0,0)$.

\begin{figure}
\includegraphics[width=1.5 in,angle=-90.]{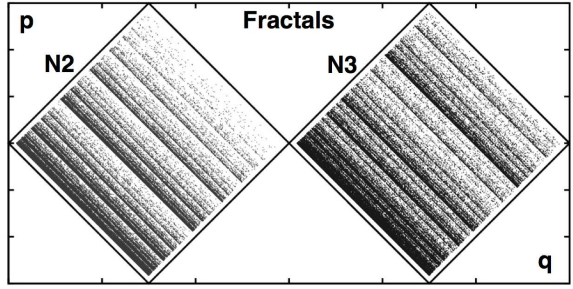}
\caption{
Fractals generated with 100,000 iterations of the N2 and N3 maps beginning at the central
point $(q,p)=(0,0)$. The information dimension estimates for the N3 map, using a variety of meshes
are all consistent with the Kaplan-Yorke prediction as well as the estimate 0.7897 based on the
mapping of regions rather than points. For the time-reversible N2 map all of these estimates differ!
See reference 38. Throughout this discussion the ``bin size'' $\delta$ is defined relative to a
$1\times 1$ unit square rather than to a $2\times 2$ area-4 diamond.
}
\end{figure}

\begin{figure}
\includegraphics[width=2.5 in,angle=-90.]{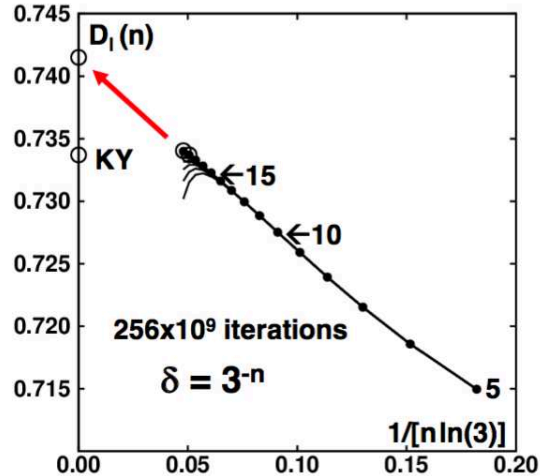}
\caption{
Dependence of the apparent information dimension for the Compressible Baker Map N2
on the bin size $\delta$.  A straight-line extrapolation from simulations with trillions of
iterations suggests an extrapolated information dimensionality of 0.741, considerably higher
than the Kaplan-Yorke estimate from the Lyapunov exponents.
}
\end{figure}

\begin{figure}
\includegraphics[width=2.5 in,angle=+90.]{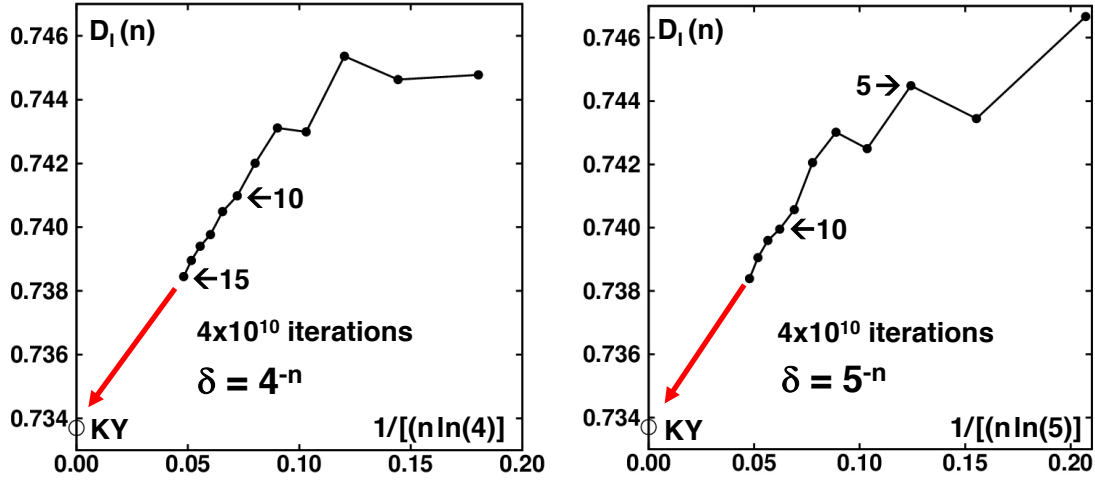}
\caption{
Apparent information dimensions in the fractal direction for the Compressible Baker Map N2. Here series of bin
sizes $\delta$ varying as inverse powers of 4 and 5 suggest good agreement with the Kaplan-Yorke estimate
of $D_{\rm KY} = 0.7337$.
}
\end{figure}

The N3 mapping was developed in an effort better to understand the information-dimension problems. Though not
time-reversible, N3 provides exactly the same set of bin probabilities as does N2 if both start with the same
uniform density.  In the N3 case the Lyapunov exponents are
$$
\lambda_1 = (2/3)\ln(3/2) + (1/3)\ln(6) = (1/3)\ln(27/2) \ ;
$$
$$
\lambda_2 = (2/3)\ln(1/3) + (1/3)\ln(1/3) = (1/3)\ln(1/27) \ [ \ {\rm N3 \ Exponents} \ ] \ .
$$
The Kaplan-Yorke prediction of the information dimension in the fractal direction, 0.7897, agrees precisely
with the region-mapping dimension shared by both maps.

Mapping a set of one million points arranged originally in a uniform grid is a convenient way to carry out
region mapping. {\bf Figure 25} compares the result of applying the N2 and N3 mapping to an initially uniform
grid of points. We choose a mesh $\delta = (1/3)^5 = 1/243$ and observe that five iterations of the maps
produce exactly the same information dimensions for both maps, as would be expected because their region
mapping probabilities are identical.  However iterations beyond the fifth show that N3's dimensionality is
unchanged at 0.7897 while the N2 dimensionality converges to 0.715, corresponding to the rightmost point in
{\bf figure 23}.

\begin{figure}
\includegraphics[width=2.5 in,angle=+90.]{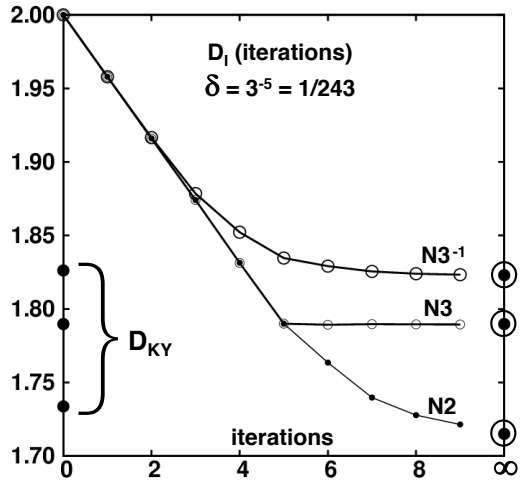}
\caption{
Information dimensions for N2, N3, and N3$^{-1}$ using 243 one-dimensional bins as functions of the number
of map iterations. We begin with a uniform distribution of one million points in the diamond-shaped domain.
$D_I$ remains the same for N2 and N3 through five iterations, after which their information dimensions differ.
The limiting values are circled. Further mesh refinements with more bins extend the agreement but in the end
the convergence remains nonuniform.  The Kaplan-Yorke conjectured dimensions are shown at the left and imply
that N3 and its inverse both obey that conjecture.  N2 may well not and furnishes an interesting example of
nonuniform convergence.
}
\end{figure}

In sum, the N3 map shows good agreement among all four methods of estimating the information dimension,
Kaplan-Yorke, region mapping, and point mapping with different values of $\delta$, suggesting that information
dimension is well understood for maps.  N2 suggests the opposite and demonstrates that the revinvestigation of
known problems with the ever-increasing computational capabilities described by Moore's Law reveals new
problems in old areas as well as access to completely new research areas.  Maps, though simple, reveal the
need for yet more investigation.

\section{60 Years' Experience with Computer Simulations}

Starting out with Kirkwood's formal 1960 approach to statistical mechanics and ending up with the 
inexpensive trillion-timestep laptop simulations of 2020 provides a rear-view-mirror look at the
effects of the computational revolution on statistical mechanics. Hours of laborious algebra and
calculus have been replaced with the need to generate, process, and understand terabytes of data.
Simulation has largely replaced theory as a means of ``understanding''. Simulations differ from
ideas. For relevance and acceptance they must be reproducible. The need for reproducibility
and cross-checking of conclusions is particularly important today when the public understanding of
``science'' has been politicized by worldwide crises, both real and imagined.

We have found that simple models capturing and detailing some aspect of reality are reliable means
of organizing, understanding, and educating. The Mayers' virial series has led to sophisticated
equation-of-state modelling. Molecular Dynamics has led from the three-body problem to simulations
with trillions of degrees of freedom. Though the research frontier is constantly on the move it
remains as tantalizing as ever despite the changing landscapes and the ever-improving tools that
we have for its exploration.

In view of the ongoing explosion of new data, results, and conjectures, fully spanning the range
from useful to useless, simple models (but not too simple) continue to provide stimulating clues
to progress in understanding, our goal through 60 years of joint explorations.  
\section{Data Availability Statement}
The data that support the findings of this study are available from the corresponding author upon reasonable request.

\pagebreak

\end{document}